\documentclass{article}
\usepackage{authblk}
\usepackage[a4paper, total={6in, 10in}]{geometry}
\usepackage{graphicx}
\usepackage{xcolor}
\usepackage{tikz}
\usepackage{amsmath}
\usepackage{amssymb}
\usepackage{hyperref}
\usepackage{cleveref}
\usepackage[backend=biber,style=phys]{biblatex}

\title{Membrane particle wrapping across methods and shapes}
\author[1,2]{Pietro Sillano}
\author[1]{Haochun Sun}
\author[1]{B. Yoldaş Cinemre}
\author[1]{Timon Idema\thanks{t.idema@tudelft.nl}}
\affil[1]{Department of Bionanoscience, Kavli Institute of Nanoscience, Delft University of Technology, Delft, The Netherlands}
\affil[2]{Molecular Dynamics Group, Groningen Biomolecular Sciences and Biotechnology Institute, University of Groningen, Groningen, The Netherlands}

\begin{document}

\maketitle
\begin{abstract}
Understanding the passive wrapping of adhesive particles by lipid membranes is a crucial step in untangling more complex endocytotic processes. However, despite the variety of computational approaches used to study passive endocytosis, a direct comparison of results across different modeling methods has never been performed. In this work, we present a comparative analysis of three distinct frameworks: numerical solutions of the axisymmetric membrane shape equation, a particle-based molecular dynamics model (MesoMem), and an energy-based triangulated surface model (FreeDTS). First, we validated our numerical solver for spherical particles against established literature and extended the implementation to include closed vesicle membranes and non-spherical particles. Next, we studied the same system with our two simulation methods, and found that the phase plots obtained from our numerical solution agree with those from our simulations. We then extended our study to anisotropic particles. By tracking engulfment pathways in MesoMem and FreeDTS, we observed a dynamic rotational transition where non-spherical particles initially bind in a horizontal (`submarine') orientation and subsequently rotate to a vertical (`rocket') orientation to complete wrapping — a behavior impossible to capture with standard axisymmetric models. Lastly, we considered the wrapping of soft vesicles with FreeDTS. Our simulations of a deformable, fixed-connectivity particle show that stronger adhesion promotes wrapping, whereas increasing membrane tension suppresses it. Complete engulfment is accompanied by a substantial reduction in particle volume, suggesting that particle compressibility provides an additional pathway for wrapping.
\end{abstract}

\section*{Statement of significance}
In order to sustain themselves and grow, cells need to internalize cargo without breaking their outer membrane. Likewise, viruses and drugs entering a cell must do so without membrane puncture. In many cases, the first step of this internalization process is the wrapping of the cargo particle by the membrane. This wrapping process has been studied extensively with different approaches, but a systematic comparison between the approaches is lacking. Here, we provide such a comparison between three complementary techniques, and build on the strengths of each technique to extend the study of wrapping to anisotropic and to soft cargo particles.

\section{Introduction}
A key component of every living cell is the lipid bilayer that forms the cell boundary, acting as a semi-permeable barrier that regulates the passage of substances in and out of the cell. The transport of small particles ($\lesssim5\;\mathrm{nm}$), such as short peptides, water molecules, or ions, can occur by simple diffusion across the membrane or be facilitated by specialized channel proteins \cite{yang_smallMolecules_transport_2015,yan_transmembrane_2025,zhang_cooperative_2015,zhang_simulation_2021}. In contrast, larger particles ($\gtrsim5\;\mathrm{nm}$) need to be wrapped by the membrane to be internalized in a process called endocytosis. Living cells have developed a wide range of pathways to internalize particles of different sizes, from clathrin-mediated routes that handle cargoes of a few tens of nanometers up to macropinocytosis, which engulfs micrometer-sized volumes~\cite{cannon_macrophage_1992,rennick_key_2021,sousa_de_almeida_understanding_2021,Zeno2021}.

Understanding the physical principles of the internalization process of nanoparticles (NPs) with sizes of $5-500\;\mathrm{nm}$ is crucial for understanding how cells can sustain themselves and grow, as well as for the development of new NPs that carry drugs into target cells with high specificity~\cite{peer_nanocarriers_2007,yusuf_nanoparticles_2023}. Experimental studies have therefore extensively explored how wrapping is influenced by fundamental physical parameters, such as particle size~\cite{rejman_size-dependent_2004,roiter_interaction_2008,le_bihan_cryo-electron_2009,lu_size_2009,he_effects_2010,wu_size-dependent_2019}, shape~\cite{chithrani_determining_2006,zuraw-weston_membrane_2021,liu_wrapping_2023,van_der_ham_particle_shape_2024}, charge \cite{harush-frenkel_targeting_2007,peetla_effect_2009,he_effects_2010,chakraborty_clathrin_2015}, and adhesion energy~\cite{groza_adhesion_2024}. Many cargos of biological or pharmaceutical interest are, moreover, deformable rather than rigid: lipid vesicles and liposomes, polymeric capsules and nano-gels, and membraneless biomolecular condensates and liquid droplets. For these, the uptake pathway depends not only on the response of the host membrane but also on how the cargo itself reshapes during wrapping~\cite{hui_nanoparticle_2020}. While these studies expand our phenomenological knowledge of the internalization process, complementary computational studies in the form of numerical models~\cite{dasguptaWrappingEllipsoidalNanoparticles2013,agudo-canalejoCriticalParticleSizes2015,agudo-canalejoStabilizationMembraneNecks2016,desernoElasticDeformationFluid2004,frey_dynamics_2019,bahrami_role_2016,decuzzi_role_nonspesific_2007,meng_size_2018} and simulations \cite{vacha_receptor-mediated_2011,chenShapedependentInternalizationKinetics2016,xiong_cooperative_2017,curtis_modeling_2015,shenMembraneWrappingEfficiency2019,shen_understanding_2018} can provide insight into the underlying mechanical processes.

Regardless of the specific details of an endocytotic pathway, the membrane always has to strongly deform to wrap the particles. Therefore, the first step in the study of these pathways is to investigate passive endocytosis, which is driven by the adhesive interactions between the particle and the membrane. From a physical point of view, passive endocytosis can be understood as a trade-off between the energetic gain of NP--membrane adhesion and the energetic cost of deforming the elastic membrane \cite{zhang_endocytosis_2015}. The membrane deformation energy can be calculated from the Helfrich Hamiltonain \cite{helfrich1973elastic}, which uses a continuous description of the membrane as a two-dimensional surface. This approximation is justified for the wrapping of NPs that are at least an order of magnitude larger than the $\approx5\;\mathrm{nm}$ bilayer thickness~\cite{manzanares_endocytosis_2020}. NP adhesion is most commonly mediated by ligand-receptor interactions, but it may also be influenced by non-specific interactions of various physical origins such as van der Waals, electrostatic, and depletion interactions~\cite{nel_understanding_2009,decuzzi_role_nonspesific_2007}. By coarse-graining these interactions into a single adhesion energy, we can treat the system as a physical continuum and solve for the equilibrium state of NP wrapping.

The continuum picture, however, describes the membrane only at mechanical equilibrium and omits the thermal fluctuations that are inherent to membranes. Molecular dynamics (MD) and Monte Carlo (MC) simulations of the membrane include thermal fluctuations, thereby expanding the phase space of possible wrapping states. Simulations with a high level of molecular detail, including full-atomistic models and coarse-grained approaches such as the MARTINI~\cite{pedersenMartini3Lipidome2025} and Cooke-Deserno~\cite{cookeCouplingLipidShape2006} 3-beads-per-lipid models, can capture the physical properties of lipid membranes. However, their complexity limits the sizes and timescales of the simulations due to extensive computational costs, while also making it difficult to interpret their results on larger-scale processes like NP wrapping. Therefore, highly coarse-grained MD and MC techniques have been developed. In these mesocale simulations, the membrane is coarse-grained into components (usually called `beads') with a small number of effective degrees of freedom. In dynamically-triangulated MC simulations, the elastic surface is discretized into such beads, while in MD approaches the membrane is assembled from interacting beads. In both cases, the local interaction rules of the beads are chosen to reproduce the elastic behavior of the membrane at large scales~\cite{munoz-basagoiti_tutorial_2025}. 

\begin{figure}[ht]
    \centering
    \includegraphics[width=\textwidth]{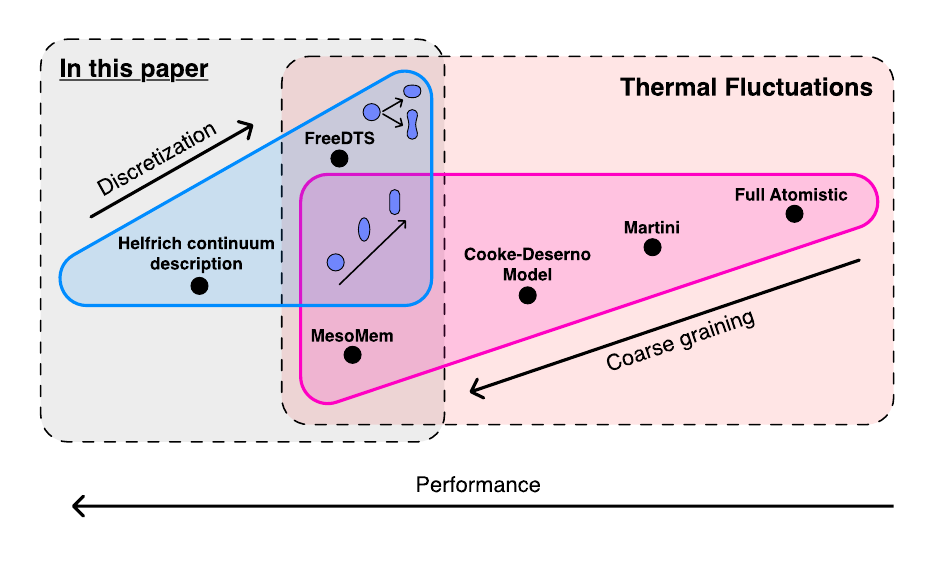}
    \caption{Schematic overview of selected computational methods for membrane simulations, contextualizing the models considered in this work. The diagonal arrows indicate two opposing methodological paradigms: the bottom-up coarse-graining approach (magenta) and the top-down discretization approach (blue). The pink shaded region encompasses models that intrinsically capture thermal fluctuations, a critical physical feature that distinguishes them from purely continuum Helfrich approaches. The bottom arrow illustrates the scaling of computational efficiency, which increases as the methods progress from highly detailed descriptions to more simplified models.}
    \label{fig:roadmap}
\end{figure}

Most existing studies rely on a single computational method to investigate the wrapping process, with little to no direct cross-comparison between methods. To bridge this gap, in this paper we aim to provide a comprehensive study of the particle wrapping process by employing three complementary computational methodologies (Fig.~\ref{fig:roadmap}): (i) numerical solution of the continuum shape equations derived from the Helfrich Hamiltonian~\cite{helfrich1973elastic}, (ii) mesoscale simulations of dynamically triangulated surfaces using FreeDTS~\cite{pezeshkianMesoscaleSimulationBiomembranes2024}, and (iii) mesoscale particle-based simulations using MesoMem~\cite{sillanoMesoMem2026}. The first two methods take a top-down approach based on the Helfrich elastic energy. The numerical solver evaluates exact axisymmetric shape equations derived from the Hamiltonian, whereas FreeDTS discretizes this energy across a fluctuating triangulated mesh. MesoMem is instead bottom-up: simple local interaction rules between particles are sufficient for the macroscopic behavior of the membrane to emerge rather than being imposed. Specifically, we study the wrapping of spherical and anisotropic particles, investigating the effects of particle adhesion, size, and aspect ratio, and membrane tension. We show that the three methods predict phase behavior that is consistent with each other, even though they are constructed with different computational approaches, while highlighting the benefits and limitations of each model. We then build on these results to study the wrapping of anisotropic hard particles and of soft elastic particles.

\section{Theoretical description of a wrapping process}

The wrapping of a particle by a lipid membrane is fundamentally driven by the competition of three energetic contributions: the adhesion energy favoring wrapping, opposed by the bending energy required to deform the membrane and by the work done against the membrane lateral tension $\sigma$, which resists the increase in membrane area needed to cover the particle.

The theoretical framework for describing this membrane deformation is provided by the Helfrich model \cite{helfrich1973elastic}. Within this continuum approach, the free energy of a fluid lipid bilayer is expressed as an integral over the local mean and Gaussian curvatures, weighted by the bending rigidity $\kappa$ and the saddle-splay modulus $\bar{\kappa}$, alongside the membrane lateral tension $\sigma$ that penalizes increases of the membrane area,
\begin{equation}
    \mathcal{F}_{\mathrm{mem}} = \int_A \left[ 2\kappa H^2 + \bar{\kappa}K + \sigma \right] \mathrm{d}A .
    \label{eq:helfrich}
\end{equation}
For wrapping that does not change the membrane topology, by the Gauss--Bonnet theorem $\int_A \bar{\kappa} K \, \mathrm{d}A$ is a constant, so the saddle-splay term does not affect the equilibrium shape and is omitted from here on.
The particle--membrane adhesion is coarse-grained into a uniform adhesion energy density $w$, so that a contact region of area $A_{\mathrm{ad}}$ contributes an energetic gain $-w A_{\mathrm{ad}}$. The total free energy of the wrapping process is therefore
\begin{equation}
    \mathcal{F} = \mathcal{F}_{\mathrm{mem}} - w A_{\mathrm{ad}} .
    \label{eq:total_energy}
\end{equation}

Throughout this paper, $\sigma$ denotes the lateral tension imposed by the surrounding membrane reservoir in the planar geometry, and the Lagrange multiplier conjugate to the total membrane area in the closed-vesicle geometry. These tensions enter the shape equations in the same way, and we therefore use the same symbol for both.

Minimizing the free energy functional~(Eq. \ref{eq:total_energy}) to determine equilibrium membrane shapes is analytically tractable only for the simplest geometries, while the full engulfment of a particle involves large, non-linear membrane deformations and the formation of a narrow membrane neck.


Our three approaches for studying membrane wrapping differ both in how the membrane is represented and in how the wrapping state is obtained: either by directly solving the shape equations, or by simulating a fluctuating membrane. The numerical solver and FreeDTS both start from the Helfrich energy, the former reducing it to exact shape equations and the latter discretizing it on a fluctuating mesh, whereas MesoMem builds a membrane with the same elastic properties from simple local rules between its constituent particles. 

\subsection{Continuum shape equations} \label{sec:numerical}
The general shape equation for a membrane vesicle has been derived from the Helfrich Hamiltonian by Ou-Yang and Helfrich~\cite{Ou-Yang1989}. The general shape equation is a nonlinear partial differential equation, which can be reduced to an ordinary differential equation for axisymmetric vesicles~\cite{seifertShapeTransformationsVesicles1991,Julicher1994}. This ordinary differential equation has been solved only for the special case that the membrane has zero surface tension and no pressure difference, which is impossible for a closed vesicle~\cite{Zhang2018}. For the onset of particle wrapping, the vesicle shape equation can be simplified by invoking a small deformation gradient approximation, as done analytically by Deserno and Bickel~\cite{deserno2003wrapping}, but this approximation breaks down at larger wrapping angles. To describe the full wrapping process, the membrane shape equation must therefore be solved numerically. The numerical solutions for axisymmetric shape equations coupled with a wrapping process have been explored extensively for approximately flat membranes \cite{desernoElasticDeformationFluid2004,zhangModelHumanImmunodeficiency2008,foretShapeEnergyMembrane2014,agudo2021particle} and spherical vesicles \cite{agudo-canalejoCriticalParticleSizes2015,agudo-canalejoStabilizationMembraneNecks2016}. 

Starting from these studies, we implement a numerical solver covering wrapping by both flat and vesicular membranes. At a fixed wrapping angle $\phi$ the membrane is divided into a bound part, which follows the shape of the wrapped particle, and whose bending and tension energy is known in closed form, and a free (unbound) part whose shape is not known a priori. We obtain the unbound shape by solving the axisymmetric Helfrich shape equation (Supplementary Note~\ref{supp-app:general_shape_equation}), parametrized by arc length and integrated with a shooting method, matching smoothly to the bound cap at the contact line and to the appropriate outer boundary: an asymptotically flat membrane in the planar case, and closure at the opposite pole in the vesicle case (Figs.~\ref{fig:phase_plot_comparison}A and~\ref{supp-fig:num_schem}).

Summing the unbound elastic energy, the analytic bound-membrane energy, and the adhesion term $-w A_{\mathrm{ad}}$ gives the total free energy $\mathcal{F}(\phi)$ of a membrane wrapping a particle at wrapping angle~$\phi$. Sweeping $\phi$ traces the wrapping energy landscape, whose minima and barriers define the equilibrium state. Representative landscapes are shown in Fig.~\ref{fig:phase_plot_comparison}B: a monotonically increasing $\mathcal{F}(\phi)$ (stable free state), a monotonically decreasing one (stable fully wrapped state), and a bistable case in which a partially wrapped minimum is separated from the free state by a barrier.

\subsection{Dynamically triangulated surfaces: FreeDTS} \label{sec:simulations_FreeDTS}
For the Monte Carlo (MC) simulations of the dynamically triangulated membrane surface we use the FreeDTS software package~\cite{pezeshkianMesoscaleSimulationBiomembranes2024}. In this mesoscopic approach, the continuous lipid bilayer is discretized into a fluid 2D network of vertices and triangular faces. The system explores its configuration space through two types of MC moves, vertex displacements and link flips \ref{supp-app:freedts}, weighted by the discretized Helfrich Hamiltonian. Unlike the numerical solver described above, which relies on the axisymmetry of the system, the dynamically triangulated surface method allows the membrane to freely break symmetry and naturally captures the thermal membrane undulations that can facilitate or hinder the wrapping transition. 

\subsection{Particle-based model: MesoMem} \label{sec:simulations_MesoMem}
For the molecular dynamics (MD) simulations, we use the particle-based MesoMem membrane model ~\cite{sillanoMesoMem2026}, implemented in the MD engine LAMMPS \cite{LAMMPS}. Unlike other coarse-grained approaches that require explicit solvent particles or multi-bead lipid representations, MesoMem uses a solvent-free, one-particle-thick framework driven by a purely additive potential. This formulation reduces the computational cost while still recovering macroscopic mechanical properties, such as the bending rigidity and tension, from local, pairwise particle interactions. Moreover, like the dynamically triangulated surface simulations, particle-based simulation allows the membrane to break symmetry and captures the thermal fluctuations of the membrane.

\bigskip
In both FreeDTS and MesoMem the membrane lateral tension is controlled by letting the projected area of the membrane fluctuate at a fixed imposed tension: through isotropic box-rescaling Monte Carlo moves in FreeDTS, and through a barostat in MesoMem. The wrapping fraction $\alpha$ is measured on the equilibrated state as the fraction of the particle surface within contact range of the membrane, averaged over configurations once $\alpha$ has reached a plateau. Further mathematical formulations, the wrapping-fraction estimators, and implementation details for the numerical solver, FreeDTS, and MesoMem are provided in the Supplementary Material~\ref{supp-app:numerical_approach}, \ref{supp-app:freedts}, and \ref{supp-app:MesoMem}, respectively.

\section{Wrapping of hard particles}

We investigate the wrapping of hard particles across a progression of increasingly complex scenarios. First, we explore the effect of the underlying membrane geometry by comparing planar membranes and spherical vesicles. Subsequently, we examine the role of particle shape by considering anisotropic particles: spherocylindrical rods are treated in the main text, while the corresponding results for ellipsoidal particles are reported in the Supplementary Material.

\subsection{Spherical particles on a planar membrane}
As the simplest and most extensively studied case \cite{desernoElasticDeformationFluid2004,agudo2021particle}, the interaction of a spherical particle with a planar membrane serves to validate our simulation methods against the numerical solutions of the axisymmetric shape equations. We focus on the dependence of the engulfment process on the adhesion strength and the applied membrane lateral tension, the two key parameters governing nanoparticle--membrane interactions. Unless stated otherwise, both simulation methods use a bending rigidity $\kappa = 20\,k_{\mathrm{B}}T$ and a particle radius $R_{\mathrm{pa}} = 5\,\ell_{0}$, where $\ell_{0}$ denotes the elementary length unit of each method (the membrane bead diameter in MesoMem, the mesh size in FreeDTS). 

\begin{figure}[ht]
    \centering
    \includegraphics[width=\textwidth]{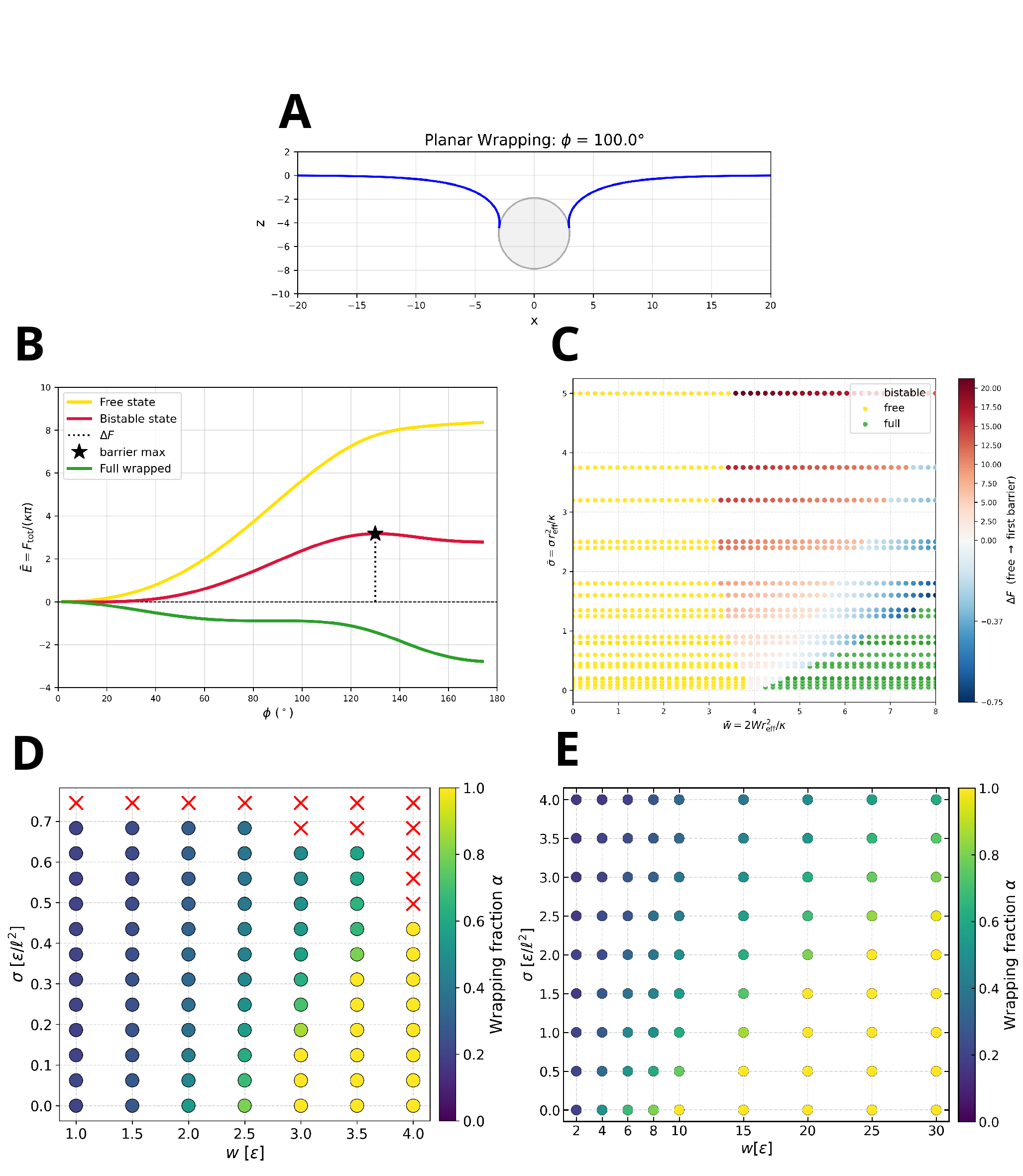}
    \caption{Wrapping of hard spherical particles. (A) Representative snapshot of a numerically found membrane shape for the wrapping of a hard particle at $\phi = 100^\circ$. Supplementary movie S1 shows the progressive increase of wrapping angle $\phi$. (B) Representative numerically obtained free energy ($\mathcal{F}$) landscapes as a function of the wrapping angle. Line colors denote the resulting equilibrium configuration: the free state (yellow), the fully wrapped state (green), and a bistable regime (red). The black star indicates the local energy maximum (transition state) in the bistable landscape. In the regime where the free state is metastable, the barrier $\Delta\mathcal{F}>0$ is measured from this maximum relative to the free minimum; where no maximum separates the two states, $\Delta\mathcal{F}<0$ is measured instead from the partially wrapped minimum and gives its depth below the free state. (C) Phase diagram derived from the energy landscapes. Scatter points are color-coded by state: solid yellow indicates the free state, solid green represents the fully wrapped state, and a gradient colormap denotes the bistable regions, within which $\Delta\mathcal{F}$ is rendered on a diverging scale whose sign change marks the loss of local stability of the free state (spinodal). (D, E) Phase diagrams showing the equilibrium wrapping fractions obtained from (D) MesoMem and (E) FreeDTS simulations, both at $\kappa = 20\,k_{\mathrm{B}}T$ and particle radius $R_{\mathrm{pa}} = 5\,\ell_0$. Each data point represents the average of three independent replicas at a given adhesion strength $w$ and membrane lateral tension $\sigma$. The color scale corresponds to the mean wrapping fraction. Supplementary movie S2 shows one wrapping trajectory for FreeDTS simulation.}
    \label{fig:phase_plot_comparison}
\end{figure}

The wrapping behavior for the numerical solution is determined from the free-energy landscape $\mathcal{F}(\phi)$, sketched for representative cases in Fig.~\ref{fig:phase_plot_comparison}B. At zero membrane lateral tension the landscape is monotonic, as established previously for this geometry \cite{desernoElasticDeformationFluid2004,deserno2003wrapping}: it either increases with $\phi$, so the only minimum is the free state $\phi = 0$, or decreases, so the only minimum is the fully wrapped state $\phi = \pi$. There is no stable partial state, and as the adhesion strength $w$ increases the equilibrium jumps discontinuously from free to fully wrapped at a single threshold.

At finite lateral tension, the contribution of the tension term grows faster than linearly with~$\phi$, curving the landscape, so that a partially wrapped minimum $0 < \phi^{\ast} < \pi$ becomes possible. Increasing $w$ then takes the system through a sequence of states. At low $w$ the free state is the only minimum. Beyond a first threshold a second, partially wrapped minimum appears while the free state is still the global minimum, so the partially wrapped state is metastable. At the binding transition the two minima are at equal energy, above it, the partially wrapped state is the global minimum, but the free state remains locally stable and is separated from the global minimum by an energy barrier. As $w$ increases further, this barrier vanishes: the free state ceases to be even locally stable. Finally, for large $w$ the wrapping angle increases until $\phi = \pi$, and the particle becomes fully wrapped.

Fig.~\ref{fig:phase_plot_comparison}C shows the wrapping phase diagram as a function of rescaled adhesion strength and membrane tension. Our numerically obtained phase diagram agrees with the analytical result of Deserno~\cite{desernoElasticDeformationFluid2004} (reproduced in Supplementary Fig.~\ref{supp-fig:rpa_plot_comparison}). The bistable region is a wedge whose apex lies on the zero-tension axis, at the finite adhesion strength where the free state transitions to the wrapped state directly, and which widens with increasing tension, opening up room for a bistable state. We augment this diagram by encoding the local shape of $\mathcal{F}(\phi)$ into each scatter point. States are color-coded in yellow for a single free minimum, green for a single wrapped minimum, and a gradient inside the bistable region, and the barrier $\Delta\mathcal{F}$ is shown on a diverging scale. Where the free state is metastable, $\Delta\mathcal{F} > 0$ is the height of the transition-state maximum above the free minimum (the black star in Fig.~\ref{fig:phase_plot_comparison}B). Where no maximum separates the two states, $\Delta\mathcal{F} < 0$ is instead measured from the partially wrapped minimum and gives its depth below the free state, so the sign change of $\Delta\mathcal{F}$ marks the spinodal, i.e., the boundary line at which the free state loses local stability.

Fig.~\ref{fig:phase_plot_comparison} also shows the equilibrium wrapping fraction~$\alpha$ obtained from MesoMem (panel~D) and FreeDTS (panel~E) on a grid of adhesion strength $w$ and membrane lateral tension $\sigma$. The comparison between the three phase diagrams below is qualitative and structural rather than quantitative: the numerical solver works in dimensionless units, whereas FreeDTS and MesoMem each use their own reduced units, as detailed in the Supplementary Material. Both simulation methods reproduce the result of the numerical solver: a free region at low $w$, a fully wrapped region at high $w$, and a band of partially wrapped states between them that broadens with tension. Two differences stand out. The first concerns the zero-tension limit, where the numerical solution admits no intermediate state and the transition is therefore discontinuous, whereas in the simulations  we find partially wrapped configurations also at zero tension, which we attribute to the presence of thermal fluctuations. The second difference is technical: MesoMem cannot sustain the highest tensions, as the membrane sheet ruptures once it is stretched beyond a threshold; the affected runs are marked with a red cross in Fig.~\ref{fig:phase_plot_comparison}D. FreeDTS is free of this limitation thus can be used to reach higher tensions, as shown in Fig.~\ref{fig:phase_plot_comparison}E.

The role of particle size is summarized in Supplementary Fig.~\ref{supp-fig:rpa_plot_comparison}. The phase boundary between the free and partially wrapped states shifts downwards monotonically with particle size, so that larger particles are wrapped at weaker adhesion. Conversely, the transition to complete wrapping is non-monotonic: intermediate-sized particles require the strongest adhesion to achieve full engulfment, whereas both smaller and larger particles become fully wrapped at lower interaction strengths. We use this survey primarily to fix the particle size for the simulations, selecting a value for which the free, partially wrapped, and fully wrapped states are all reachable; the same criterion guides the choice of a representative non-zero lateral tension.

\subsection{Spherical particles on a vesicle}

In the planar membrane approximation employed in the previous section we assume that the engulfed particle is significantly smaller than the surrounding membrane. However, when the particle size is of the same order as the vesicle itself, this approximation breaks down, necessitating a direct model of the closed spherical vesicle. To capture this regime, we consider a closed vesicle with both its membrane area and its enclosed volume held fixed, and a rigid spherical particle adhering to its outer surface.

Following the work of Agudo-Canalejo and Lipowsky~\cite{agudo-canalejoCriticalParticleSizes2015}, we implement a numerical solver of the axisymmetric shape equations for a spherical particle interacting with a closed vesicle. This system of ordinary differential equations (ODEs) is widely used \cite{seifertShapeTransformationsVesicles1991}, but it is considerably more prone to numerical instabilities than the planar equivalent, and open-source available implementations of the full solving protocol remain scarce. We therefore provide a freely available implementation in our GitLab repository \url{https://gitlab.tudelft.nl/idema-group/particlewrapping.git}. Stable convergence is achieved with a multiple-shooting method combined with a south pole expansion that regularizes the coordinate divergences, following Christ et al.~\cite{christActiveShapeOscillations2021} and Frey et al.~\cite{freyMembraneAreaGain2022}, together with a systematic refinement of the initial parameter guesses. Supplementary movie S3 shows the progressive increase of wrapping angle $\phi$. These numerical strategies are detailed in Supplementary Note~\ref{supp-app:multi_shooting}.

Sweeping the spontaneous curvature $m$ and the relative particle size $R_{\mathrm{pa}}/R_{\mathrm{ves}}$, our solver recovers the four states reported by Agudo-Canalejo and Lipowsky~\cite{agudo-canalejoCriticalParticleSizes2015}: a stable free state, a stable partially wrapped state, a stable completely engulfed state, and a bistable regime in which the free and the completely engulfed state coexist. The reproduced state diagram, together with a representative free-energy landscape for each state, is shown in Supplementary Fig.~\ref{supp-fig:vesicle_validation}.

\subsection{Anisotropic particles}
Having verified the wrapping behavior of spherical particles, we next consider prolate elliptical and spherocylindrical particles as a minimal anisotropic extension. Unlike a sphere, a prolate particle combines a lower-curvature sidewall with higher-curvature end caps and introduces particle orientation as an additional degree of freedom. We report on our results on the spherocylindrical particles here, and show those of the elliptical particles in section~\ref{supp-app:ellipsoids} of the Supplemental Material.

We consider spherocylindrical particles with radius~$r_{\text{rod}}$ and total length~$h$, so that the particle has aspect ratio $\mathrm{AR}=h/(2r_{\text{rod}})$ (Fig.~\ref{fig:rod_aspect_ratio}A). With this definition, $\mathrm{AR}=1$ corresponds to a sphere, and larger values correspond to more elongated particles. We first characterize the axisymmetric wrapping process using the numerical method and then turn to FreeDTS and MesoMem simulations to determine how rotational freedom affects the wrapping pathway.

\subsubsection{Axisymmetric wrapping of spherocylindrical particles}

\begin{figure}[ht]
    \centering
    \includegraphics[width=\textwidth]{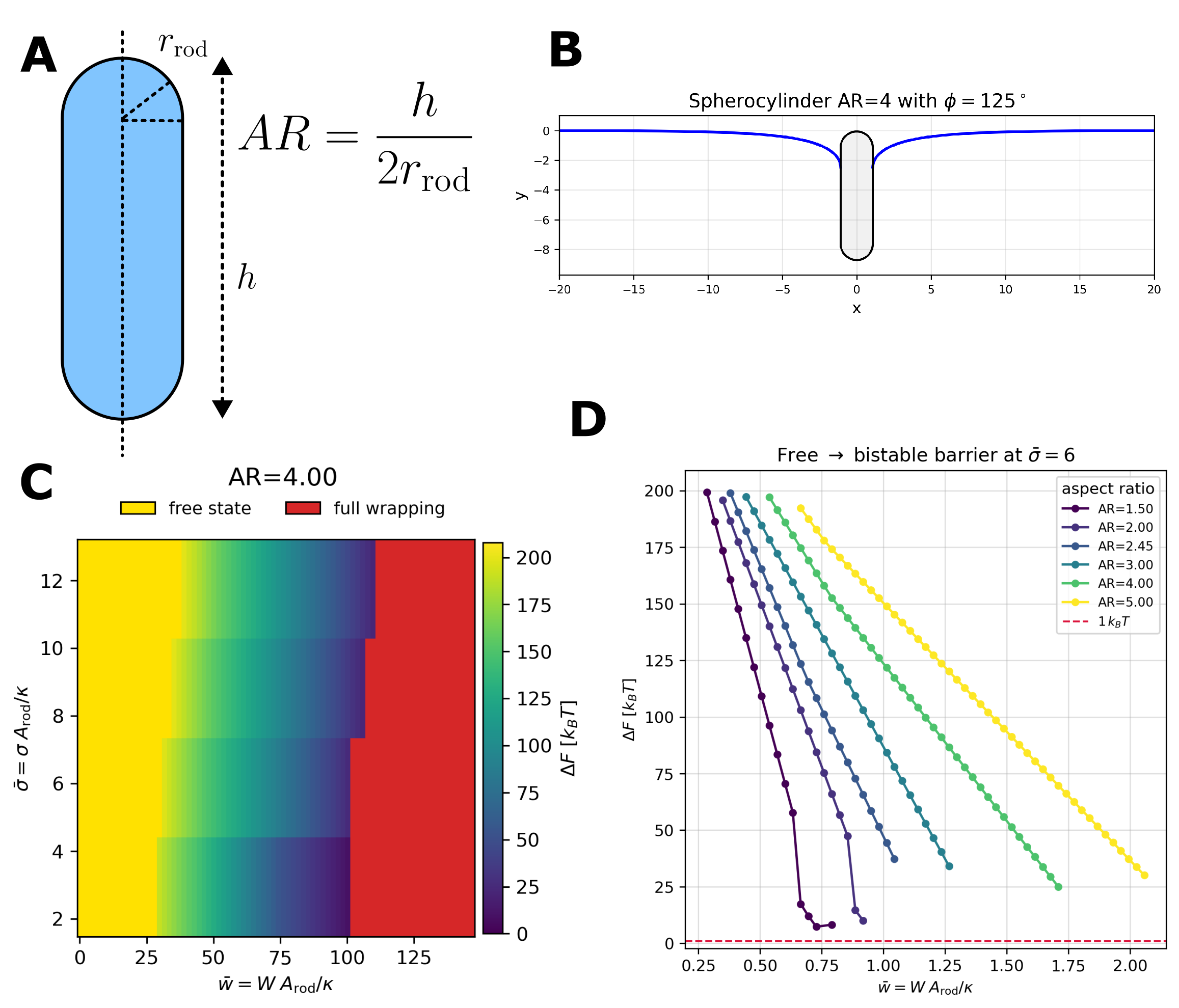}
    \caption{Wrapping of hard spherocylindrical particles with constrained orientation. (A) Schematic of the spherocylindrical particle, defining the cap radius $r_{\text{rod}}$, the cylinder length $h$, and the aspect ratio $\text{AR}=h/(2r_{\text{rod}})$. (B) Snapshot of a single numerical solution of the wrapping of a spherocylindrical particle with aspect ratio $4$ at wrapping angle $\phi = 125^\circ$. Supplementary movie S4 shows the progressive wrapping increasing $\phi$. (C) Free $\rightarrow$ bistable energy barrier $\Delta\mathcal{F}$ as a function of adhesion strength $\bar{w}$ and membrane lateral tension $\bar{\sigma}$, for a spherocylinder shape with aspect ratio $4.0$. Yellow marks the free (unwrapped) state, red the fully wrapped state, and the viridis gradient shows the barrier magnitude within the bistable region. Increasing aspect ratio shifts both the free $\rightarrow$ bistable and bistable $\rightarrow$ full boundaries to higher $\bar{w}$ and widens the bistable band. (D) Energy barrier value at fixed lateral tension $\bar{\sigma}=6$ for different aspect ratios spanning $1.5-5.0$. The red dashed line shows the thermal energy $k_{\mathrm{B}}T$.}
    \label{fig:rod_aspect_ratio}
\end{figure}

We extended the axisymmetric numerical shape solver to spherocylindrical particles with different aspect ratios. Axial symmetry constrains the particle's long axis to remain aligned with the normal to the far-field membrane, corresponding to the rocket-like configuration shown in Fig.~\ref{fig:rod_aspect_ratio}B. The numerically obtained wrapping energy landscapes fall into the same three types already illustrated for spherical particles in Fig.~\ref{fig:phase_plot_comparison}B: a monotonically increasing landscape with a stable unwrapped state, a monotonically decreasing one with a stable fully wrapped state, and a bistable regime in which a partially wrapped minimum is separated from the unwrapped state by an energy barrier.
The phase plot in Fig.~\ref{fig:rod_aspect_ratio}C shows that, as expected, increasing the surface tension shifts both wrapping boundaries towards higher adhesion strengths. At fixed membrane lateral tension, the height of the wrapping energy barrier decreases with increasing adhesion strength $w$ but generally remains well above $k_{\mathrm B}T$ (Fig.~\ref{fig:rod_aspect_ratio}D). If we increase the aspect ratio at fixed total area, the position of the barrier shifts to higher adhesion strengths. Thus, within the constrained rocket-like configuration, more elongated rods require stronger adhesion to achieve complete wrapping and are more likely to become kinetically trapped in a partially wrapped state.

\subsubsection{Simulations of rod wrapping}

The axisymmetric solver restricts the particle long axis to remain parallel to the far-field membrane normal and therefore describes only an orientationally constrained, rocket-like (vertical orientation) process. It cannot capture side-on attachment or particle reorientation. However, coarse-grained simulations have demonstrated the influence of particle shape on receptor-mediated uptake~\cite{vacha_receptor-mediated_2011} and identified a laying-down-then-standing-up sequence for initially upright spherocylinders~\cite{huang2013role}. Likewise, experiments on the wrapping of spherocylindrical rods by giant unilamellar vesicles (GUVs) have identified the side-first pathway to be the dominant way~\cite{van_der_ham_particle_shape_2024}. We therefore turn to our simulations to resolve the actual wrapping process of an anisotropic particle.

\begin{figure}[ht]
    \centering
    \includegraphics[width=0.95\textwidth]{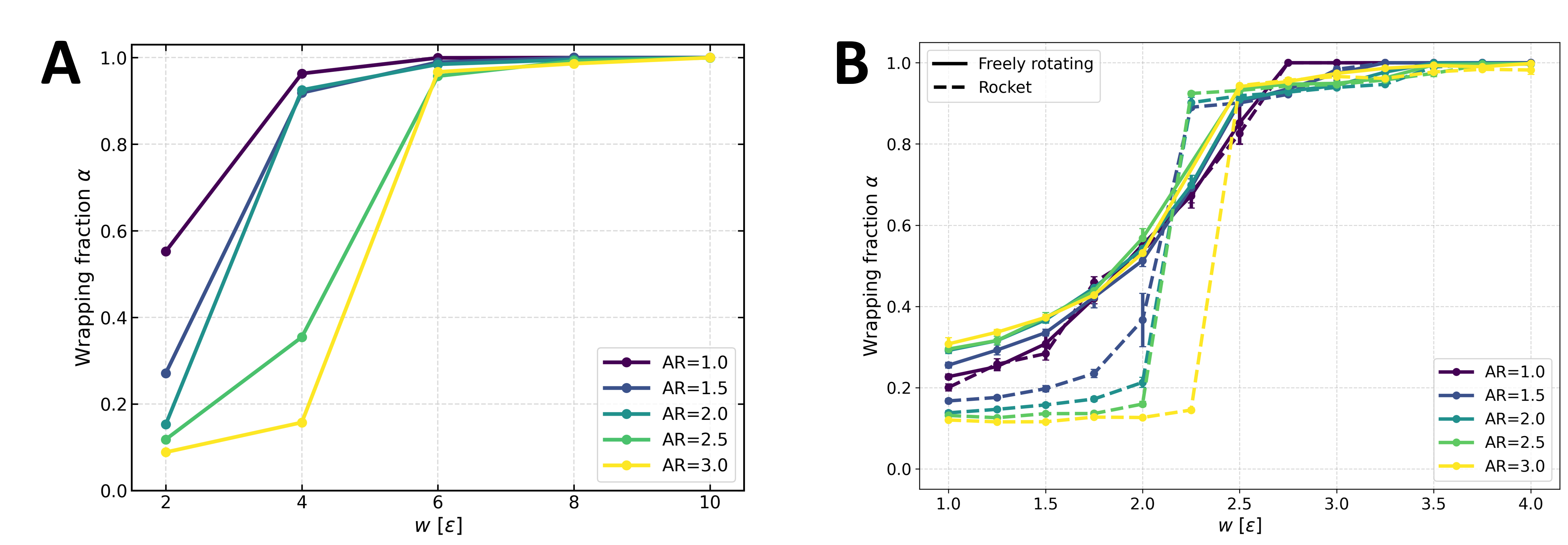}
    \caption{Wrapping fraction $\alpha$ versus interaction strength $w$ for equal-area rods at zero membrane tension. Colors indicate the particle aspect ratio. Rods are nonrotating with their long axes perpendicular to the initial membrane plane in (A) FreeDTS and in the dashed curves of (B) MesoMem; solid curves in (B) correspond to freely rotating rods.}
    \label{fig:rod_constrained_wrapping}
\end{figure}

\paragraph{Constrained rocket-like wrapping}
To connect with the axisymmetric calculations, we first constrain the rod axis perpendicular to the planar membrane. The FreeDTS and MesoMem simulations are performed at zero membrane tension, with the particle surface area fixed, while the aspect ratio is varied (Fig.~\ref{fig:rod_constrained_wrapping}). Both methods show that more elongated rods generally require stronger interactions to reach full wrapping, while all investigated shapes approach complete coverage at sufficiently strong adhesion.

This trend is expected based on the geometry of equal-area spherocylinders: increasing the elongation reduces the particle radius and increases the end-cap curvature, raising the local bending-energy density required for membrane conformity.

The high-curvature end cap thus acts as a geometrical bottleneck of the constrained rocket-like pathway. To test whether lateral contact and particle rotation can bypass this bottleneck, we next release the orientational constraint and examine freely rotating rods under the same zero-tension conditions.

\begin{figure}[ht]
    \centering
    \includegraphics[width=0.95\textwidth]{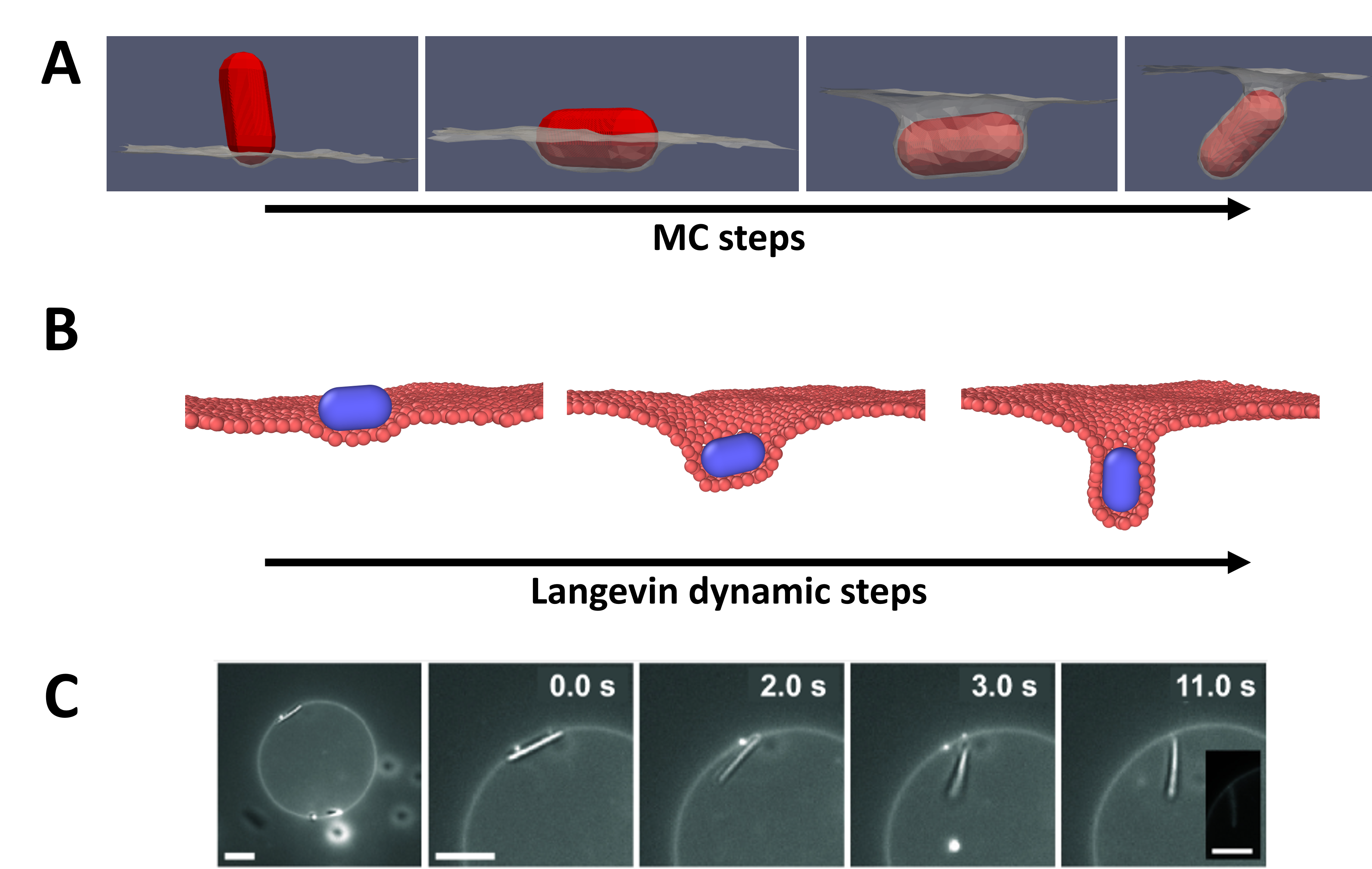}
    \caption{
    Wrapping of hard spherocylindrical particles with orientational freedom. 
    (A) FreeDTS snapshots of the submarine-to-rocket reorientation described in the text. Supplementary movie S5 shows the corresponding trajectory.
    (B) Corresponding MesoMem snapshots showing the progressive formation of the membrane invagination. Supplementary movie S6 shows the corresponding trajectory.
    (C) Experimental side-first engulfment of a rod by a giant unilamellar vesicle, adapted from Fig.~2a of \cite{van_der_ham_particle_shape_2024} under the terms of the CC BY 4.0 license. The white scale bars in (C) represent $5\,\mu\mathrm{m}$.}
    \label{fig:rod_dynamic}
\end{figure}

\paragraph{Rotation-mediated wrapping of freely rotating rods}
In both our simulation methods, we found that allowing the rods to rotate as they are wrapped, causes them to adopt a laying-down–then-standing-up pathway similar to the one found by Huang et al.~\cite{huang2013role}. In the representative FreeDTS trajectory shown in Fig.~\ref{fig:rod_dynamic}A, the rod initially contacts the membrane through its lower end cap. As contact extends onto the cylindrical sidewall, it rotates towards the membrane plane and adopts a side-on, submarine-like configuration. Wrapping then proceeds while the rod remains approximately parallel to the membrane. At higher wrapping fractions, the rod gradually tilts away from the membrane plane, returning towards a more upright, rocket-like orientation. The same qualitative submarine-to-rocket reorientation is reproduced in MesoMem simulations (Fig.~\ref{fig:rod_dynamic}B) and is consistent with the experimental side-first pathway observed for rods interacting with giant unilamellar vesicles~\cite{van_der_ham_particle_shape_2024} (Fig.~\ref{fig:rod_dynamic}C).

The MesoMem results also show that allowing the rods to rotate, generally increases their wrapping fraction at weak and intermediate interaction strengths, particularly for more elongated rods (Fig.~\ref{fig:rod_constrained_wrapping}B). This difference diminishes at strong adhesion, where both constrained and freely rotating particles approach complete coverage. The constrained rocket-like pathway thus overestimates the interaction strength required for full wrapping, especially for elongated rods.

\begin{figure}[ht]
    \centering
    \includegraphics[width=0.95\textwidth]{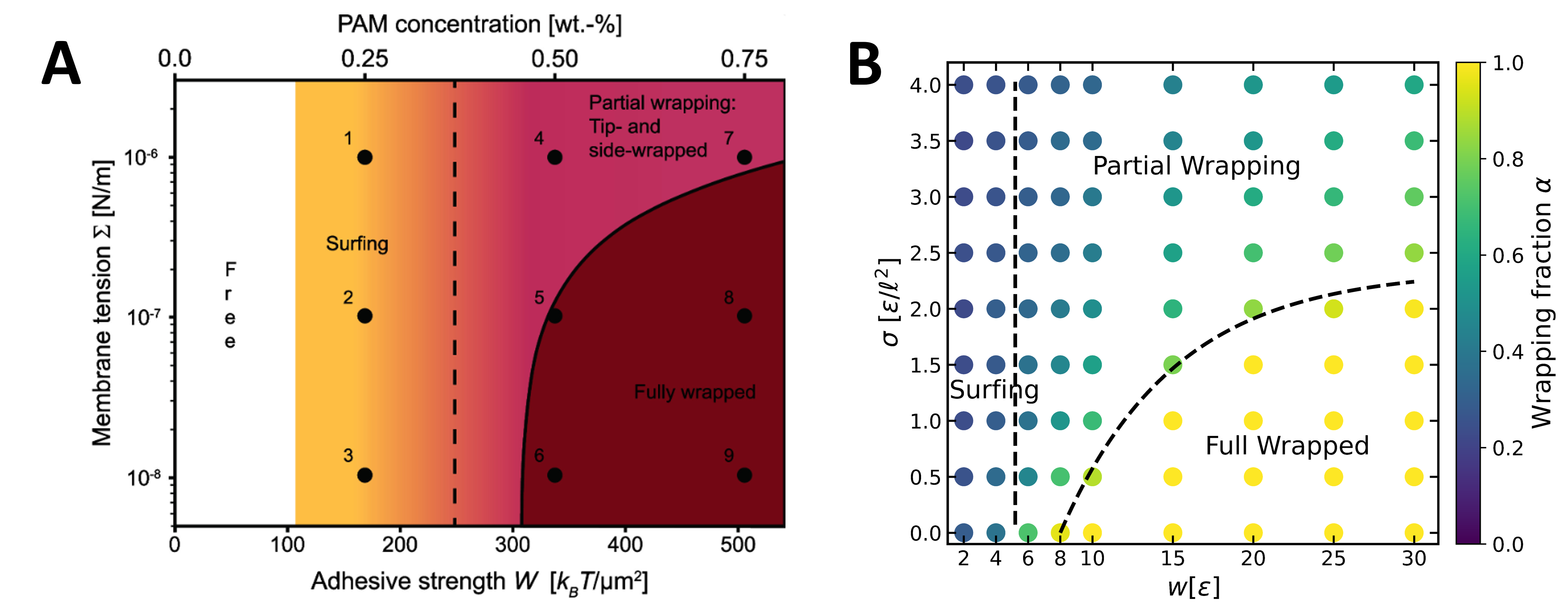}
    \caption{
    Comparison of experimental and simulated wrapping results for rod-shaped particles. (A) Phase plot combining experimentally observed wrapping states of rods interacting with giant unilamellar vesicles and theoretically predicted state boundaries as functions of the adhesion free-energy density $W$ and membrane tension $\Sigma$. Adapted from Fig.~4b of Van der Ham et al.~\cite{van_der_ham_particle_shape_2024} under the terms of the CC BY 4.0 license. (B) Final wrapping fraction $\alpha$ obtained from FreeDTS simulations of a rod interacting with an initially planar membrane as a function of the adhesion strength $w$ and reduced membrane lateral tension $\sigma$.}
    \label{fig:rod_adhesion_tension}
\end{figure}

To construct a phase plot of the wrapping of our spherocylindrical rod, we classify membrane-associated configurations with $\alpha<0.2$ as surfing-like, those with $0.2\leq\alpha<0.9$ as partially wrapped, and those with $\alpha\geq0.9$ as fully wrapped. In Fig.~\ref{fig:rod_adhesion_tension} we compare resulting phase plot from our FreeDTS simulations with the experimental results of Van der Ham et al.~\cite{van_der_ham_particle_shape_2024}. Despite the differences between the two systems, both phase plots exhibit the same qualitative organization of wrapping states: increasing adhesion promotes wrapping, whereas increasing membrane tension suppresses it and shifts the transition towards stronger adhesion. Thus, low adhesion favors surfing-like states, intermediate conditions produce partial wrapping, and strong adhesion at low tension leads to complete wrapping. This common trend reflects the competition between the adhesion energy gained from membrane-particle contact and the elastic and tensional costs of membrane deformation.

Comparing the simulated rod phase plot (Fig.~\ref{fig:rod_adhesion_tension}B) with the corresponding spherical particle phase plot (Fig.~\ref{fig:phase_plot_comparison}D) shows that the boundary between partially and fully wrapped states is approximately linear for the sphere but distinctly curved for the rod. For a sphere, with uniform curvature, the energetic costs of progressing towards complete wrapping vary relatively smoothly. For a rod, the partially wrapped state can evolve through lateral contact, end-cap wrapping and particle reorientation. These changes modify the bending cost, adhered area and recruited membrane area along the transition, resulting in a nonlinear relation between the membrane tension and the adhesion strength required for full wrapping.

\section{Wrapping of a deformable particle}
The wrapping of hard particles provides a controlled setting for investigating membrane mechanics~\cite{van_der_ham_particle_shape_2024,vanderwelLipidMembranemediatedAttraction2016}. In practice, however, the particles are often soft. Deformable particles are relevant to a broad range of biological and synthetic cargoes, including liposomes, vesicles, polymeric capsules, and nanogels. We therefore also consider the case of a deformable, initially spherical vesicle interacting with an initially planar membrane.  In contrast to a hard particle, for which all shape deformation is limited to the wrapping membrane, a deformable particle can redistribute the deformation between the host membrane and its own surface. The wrapping state is therefore controlled not only by adhesion, membrane tension, and particle size, but also by the mechanical properties and geometric constraints of the particle~\cite{yiCellularUptakeElastic2011, debnath2025wrapping}.

Particle deformability can have competing effects on membrane wrapping. Deformation may facilitate the formation of an initial contact region by allowing the particle to flatten against the host membrane. At larger wrapping fractions, however, part of the adhesion-energy gain must compensate for deformation of the particle itself. The resulting flattened or nonspherical configurations may also generate highly curved regions near the contact line, thereby stabilizing partially wrapped states and increasing the adhesion required for complete engulfment~\cite{yiCellularUptakeElastic2011,midyaMembraneMediatedInteractionsNonspherical2023,sahuWrappingNonsphericalVesicles2025}. Recent experiments on vesicle-vesicle engulfment have similarly shown that deformability can suppress the transition from shallow or full wrapping to complete internalization when the vesicle size is comparable to the bendocapillary length $\lambda = \sqrt{\kappa/\sigma}$~\cite{vanderhamShallowFullWrapping2025}.

In our simulations, the particle is represented by a closed triangulated membrane with bending rigidity~$\kappa_{\mathrm{NP}}$, while the initially planar host membrane has bending rigidity~$\kappa_{\mathrm{M}}$ and frame tension~$\sigma$. The two membranes interact through an adhesion potential of strength $w$. The particle area and reduced volume are controlled by harmonic constraints. Although coupled continuum formulations for deformable vesicles have previously been developed~\cite{yiCellularUptakeElastic2011}, they require the simultaneous determination of the host-membrane shape, the particle shape, and the moving contact boundary. This coupled free-boundary problem is not accessible to the fixed-particle numerical solver used above. We therefore investigate deformable-particle wrapping using FreeDTS simulation. 

\paragraph{Deformation during a representative wrapping trajectory}

\begin{figure}[ht]
    \centering
    \includegraphics[width=0.95\textwidth]{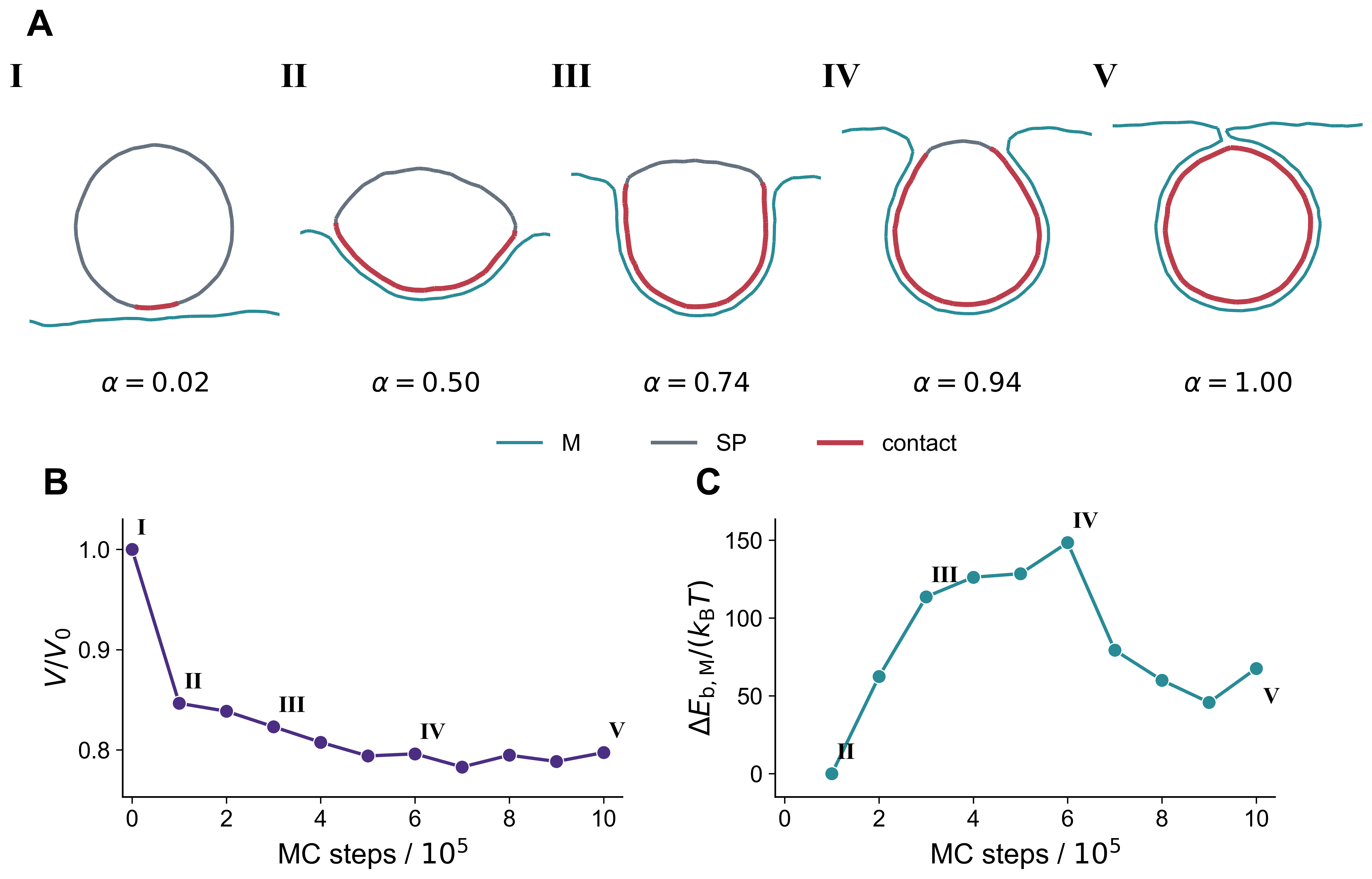}
    \caption{Representative wrapping trajectory of a soft particle at $w=6\,k_{\mathrm B}T$, $\sigma=1\,k_{\mathrm B}T/\ell_0^2$, and $\kappa_{\mathrm{M}}=\kappa_{\mathrm{SP}}=20\,k_{\mathrm B}T$. (A) Mesh cross-sections at $0$, $1$, $3$, $6$, and $10\times10^5$ MC steps (I-V). The wrapping fraction $\alpha$ is indicated below each section. Teal and gray denote the membrane (M) and soft particle (SP), respectively; red marks the contact region. (B) Change of the relative particle volume $V/V_0$, where $V_0$ is the initial volume, as the particle gets wrapped. (C) Change in planar-membrane bending energy relative to the configuration at $10^5$ MC steps.}
    \label{fig:single trajectory}
\end{figure}

Fig.~\ref{fig:single trajectory}A shows a representative trajectory of soft particle wrapping. As the wrapping fraction increases from $0$ to $1$, the soft particle volume decreases from its initial volume $V_0$ to approximately $0.8V_0$ (Fig.~\ref{fig:single trajectory}B). The cross-sections show a flattened particle during early partial wrapping, a transient droplet-like profile at deeper wrapping, and a rounder profile after full coverage. The membrane bending energy reaches a maximum near complete wrapping and decreases as full coverage is reached (Fig.~\ref{fig:single trajectory}C). 

\paragraph{Adhesion--tension control of soft-particle wrapping}

\begin{figure}[ht]
    \centering
    \includegraphics[width=0.8\textwidth]{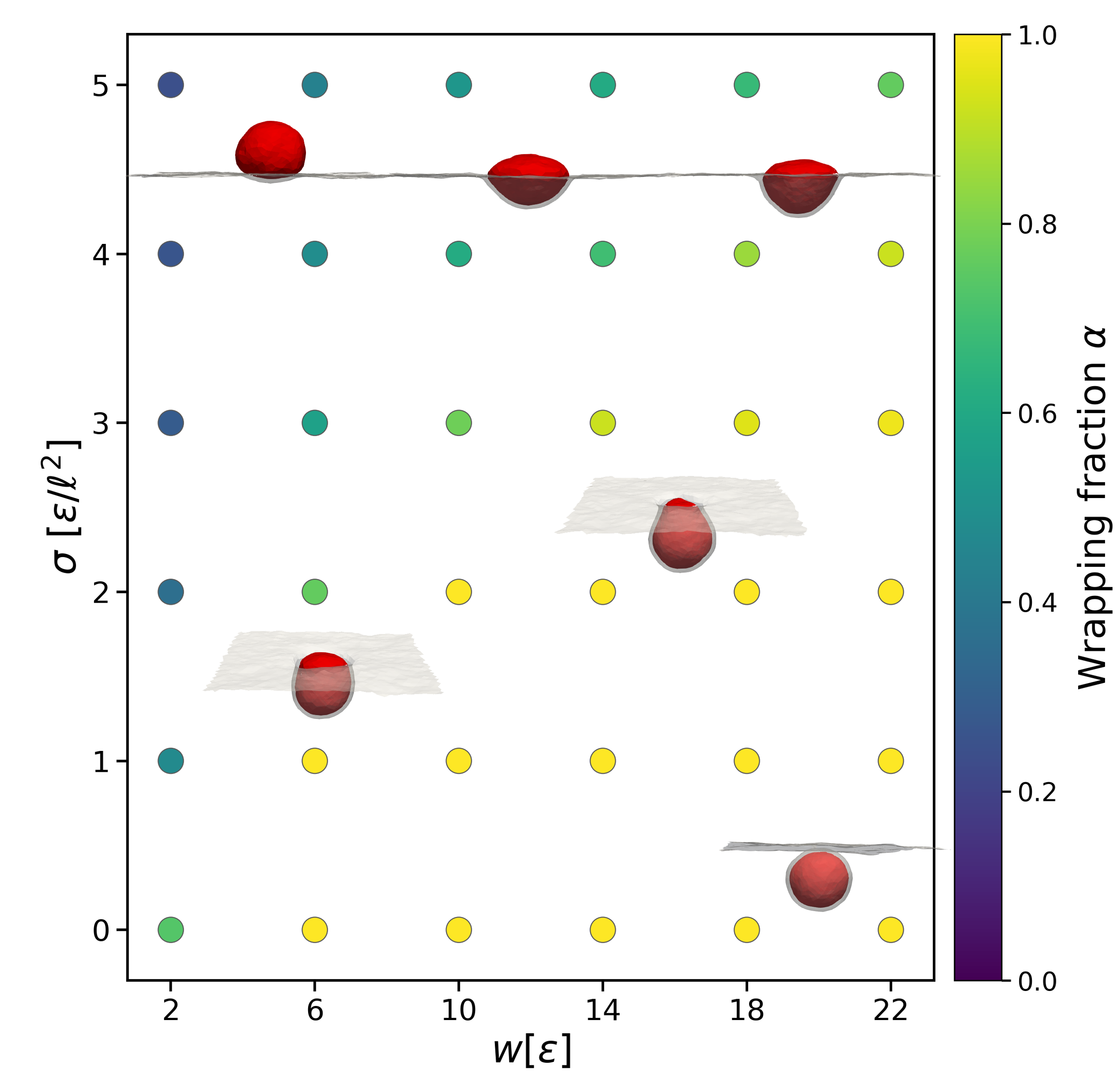}
    \caption{Phase diagram of the soft particle interacting with a planar membrane at $\kappa_{\mathrm{M}}=\kappa_{\mathrm{SP}}=20\,k_{\mathrm{B}}T$. Each symbol corresponds to a separate simulation performed at the indicated adhesion strength $w$ and frame tension $\sigma$, with color denoting the final-state wrapping fraction. Representative configurations are shown in red (soft particle) and gray (planar membrane).}
    \label{fig:soft_adhesion_tension}
\end{figure}

To assess how the wrapping of a soft particle depends on adhesion and tension, we systematically vary the adhesion strength~$w$ and the membrane tension~$\sigma$ over a discrete parameter grid, while fixing the bending rigidities at $\kappa_{\mathrm{M}} = \kappa_{\mathrm{SP}} = 20\,k_{\mathrm{B}}T$. Separate simulations were performed over a discrete grid of $w$ and $\sigma$. For each parameter pair, the wrapping fraction $\phi_{\mathrm{w}}$ was calculated from the final configuration after $10^{6}$ Monte Carlo steps. The resulting phase plot is shown in Fig.~\ref{fig:soft_adhesion_tension}.

The phase plot reveals the opposing effects of adhesion and frame tension. At a given frame tension, increasing the adhesion strength drives the system from weak or partial wrapping towards near-complete wrapping. Conversely, increasing the frame tension favors less fully wrapped configurations and shifts the partial-to-full-wrapping crossover towards stronger adhesion. Consequently, moderate-to-strong adhesion is sufficient to produce near-complete wrapping in the low-tension regime, whereas an extended region of partially wrapped states emerges at intermediate tensions. In the high-tension regime, complete wrapping is not observed over the investigated adhesion range, even under strong adhesion. While the trends for wrapping soft particles are thus the same as for hard particles, partially wrapped soft particles assume distinctly different shapes. Initial wrapping is easier, but full wrapping harder for soft particles compared to hard ones.



\section{Conclusion}
We have studied membrane wrapping of particles with three methods built on very different foundations: numerical solutions of exact axisymmetric shape equations, energy-based simulations using a dynamically triangulated surface model (FreeDTS), and force-based simulations using a particle-based mesoscale model (MesoMem). For spherical particles on a planar membrane the three methods predict phase diagrams which agree: a free state at low adhesion, a fully wrapped state at high adhesion, and a partially wrapped band in between that widens with membrane tension. That agreement is not trivial: unlike the other two methods, the particle-based MesoMem simulation is not based on the Helfrich Hamiltonian; instead, its elasticity emerges from pairwise rules between beads, and it still reproduces the phase boundary computed by solving the shape equations. The main difference is at zero tension, where the zero-temperature continuum solution does not allow for stable partial wrapping, but both simulations (which opereate at nonzero temperature) find it.

For the spherocylinder, the numerical solver can only describe the tip-first rocket pathway due to the axisymmetry constraint. Released from that constraint, rods in both FreeDTS and MesoMem bind in or turn to a side-on, submarine-like orientation and only reorient into the rocket configuration at the final stages of wrapping. Unconstrained rods reach larger wrapping fractions at weak and intermediate adhesion than their constrained counterparts, with the gap largest for the most elongated rods. 

For a deformable particle, wrapping is coupled to particle deformation in addition to membrane deformation. Adhesion promotes wrapping and membrane tension opposes it. The nonmonotonic evolution of the planar-membrane bending energy further shows that the energetic cost of wrapping is largest at an intermediate stage and is partly released during final closure. Soft particle wrapping therefore cannot be characterized by the wrapping fraction alone: the same coverage can correspond to different particle volume and deformation state.

The numerical solver returns energy landscapes, barriers, and spinodals at negligible computational cost, and should be the reference whenever the problem is genuinely axisymmetric. It cannot handle orientational degrees of freedom or symmetry breaking, and it cannot say anything about kinetics or trapping. FreeDTS can break symmetry freely, reaches high tensions without difficulty, and handles a deformable particle with controlled area and volume, at the price of Monte Carlo dynamics that carry no physical timescale and do not scale well with increasing number of vertices. The MesoMem bottom-up implementation results in a membrane whose properties are not imposed but emerge. It can deal with asymmetric systems with care of the upper tension threshold, and because it is a molecular dynamics simulation, can be used to study the evolution of membrane processes. Several extensions follow directly. MesoMem and FreeDTS allow the study of patchy and patterned particles, and thus to superselectivity, where uptake depends sharply on receptor density rather than on total binding energy, as well as multiple particles.

\section*{Code and data availability}
The full codebase used for simulations and analysis is available on GitLab at \url{https://gitlab.tudelft.nl/idema-group/particlewrapping.git}. The data that supports the findings of this article are openly available
at \cite{4TUdata}.

\section*{Author contributions}
PS and TI conceived the original idea. PS and HS developed the numerical method. PS, HS and BYC designed and performed the simulations, analyzed the data, and wrote the initial manuscript. TI acquired funding, supervised, provided critical feedback, and helped shape the final version of the manuscript.

\section*{Declaration of interests}
The authors declare no competing interests.

\section*{Acknowledgments}
We thank Felix Frey and Jaime Agudo-Canalejo for their advice on the numerical solver procedure, and Stefano Oberto for designing the diagram in Fig.~\ref{fig:roadmap}.

This research is supported by the `BaSyC - Building a Synthetic Cell' Gravitation grant (024.003.019) and the `EVOLF - Evolving life from non-life' Summit grant (SUMMIT.1.004), both of the Netherlands Ministry of Education, Culture and Science (OCW) and the Netherlands Organisation for Scientific Research (NWO).

The authors acknowledge the use of computational resources of the DelftBlue supercomputer, provided by Delft High Performance Computing Centre.

\section*{Declaration of generative AI and AI-assisted technologies in the writing process}
During the preparation of this work the authors used ClaudeAI in order to improve readability and flow of the text. After using this tool/service, the author(s) reviewed and edited the content as needed and take full responsibility for the content of the publication.

\printbibliography

\clearpage
\setcounter{section}{0}
\setcounter{figure}{0}
\setcounter{table}{0}
\setcounter{equation}{0}
\renewcommand{\thesection}{S\arabic{section}}
\renewcommand{\thefigure}{S\arabic{figure}}
\renewcommand{\thetable}{S\arabic{table}}
\renewcommand{\theequation}{S\arabic{equation}}
\renewcommand{\theHsection}{S.\arabic{section}}
\renewcommand{\theHfigure}{S.\arabic{figure}}
\renewcommand{\theHtable}{S.\arabic{table}}
\renewcommand{\theHequation}{S.\arabic{equation}}

\begin{center}
    {\LARGE Supplementary Information\par}
\end{center}

\begin{refsection}


\section{Supplementary movies}

\paragraph{Movie S1. Wrapping of a spherical particle with numerical solver.}
Spherical particle wrapped by a planar membrane with particle radius $R_{pa} = 3.0\,\ell$ and imposed membrane tension $\Sigma = 0.1\,\kappa/\ell^{2}$. Each frame is the axisymmetric membrane profile solution of the axissymmetric shape equation (shooting method) for a \emph{prescribed} contact angle $\theta$.

\paragraph{Movie S2. Wrapping of a spherical particle with FreeDTS.}
FreeDTS simulation showing the progressive wrapping of a rigid spherical particle by an initially planar membrane. 

\paragraph{Movie S3. Vesicle wrapping with numerical solver.}
Spherical particle engulfed by a vesicle. $R_{pa} = 4.5$, $R_{ves} = 30$, $w = 2$, $m = 0$. Each frame is the axisymmetric membrane profile solution of the axissymmetric shape equation (multiple-shooting method) for a \emph{prescribed} contact angle $\theta$.

\paragraph{Movie S4. Rod wrapping with numerical solver.}
Spherocylinder wrapped by a planar membrane with aspect ratio $\mathrm{AR} = 4.0$ and imposed membrane tension $\Sigma = 0.1\,\kappa/\ell^{2}$. Each frame is the axisymmetric membrane profile solution of the axissymmetric shape equation (shooting method) for a \emph{prescribed} contact angle $\theta$.

\paragraph{Movie S5. Rotation-mediated wrapping of a rigid rod with FreeDTS.}
FreeDTS simulation of a freely rotating spherocylindrical particle interacting with an initially planar membrane. The initially upright rod first contacts the membrane through an end cap, then rotates towards the membrane plane and establishes lateral contact.

\paragraph{Movie S6. Rotation-mediated wrapping of a rigid rod with MesoMem}
MesoMem simulation of a freely rotating spherocylindrical particle interacting with an initially planar membrane.

\paragraph{Movie S7. Wrapping of a deformable particle.}
FreeDTS simulation showing the wrapping of an initially spherical soft particle by an initially planar membrane.

\section{Numerical approach}
\label{supp-app:numerical_approach}
Wrapping is governed by a balance between the adhesion energy, which favours contact, and the bending and tension costs of deforming the membrane. At a fixed wrapping angle $\phi$ we split the total free energy into an unbound contribution, a bound contribution, and the adhesion energy:
\begin{equation}
    F_{\mathrm{tot}}(\phi)
    =
    F_{\mathrm{me}}^{\mathrm{un}}(\phi)
    +
    F_{\mathrm{me}}^{\mathrm{bo}}(\phi)
    +
    F_{\mathrm{ad}}(\phi).
    \label{supp-eq:total_energy}
\end{equation}
$F_{\mathrm{me}}^{\mathrm{bo}}(\phi)$ is easily calculable because the membrane follows the particle contour, whereas $F_{\mathrm{me}}^{\mathrm{un}}(\phi)$ requires solving the shape equation \cite{lipowskyStructureDynamicsMembranes1995,desernoElasticDeformationFluid2004}. Evaluating $F_{\mathrm{tot}}$ over a range of wrapping angles then yields the engulfment energy landscape, such as the ones shown in Fig.~\ref{fig:phase_plot_comparison}(B).

In the planar geometry the bound membrane follows the surface of a spherical particle, forming a cap of wrapping angle $\phi$. Measured relative to a flat patch of the same projected area, its free energy is
\begin{equation}
    F_{\mathrm{me}}^{\mathrm{bo}}(\phi)
    =
    4\pi\kappa
    \left(1+2mR_{\mathrm{pa}}\right)
    (1-\cos\phi)
    +
    \sigma\pi R_{\mathrm{pa}}^2
    (1-\cos\phi)^2,
    \label{supp-eq:bound_energy}
\end{equation}
and the adhesion contribution is
\begin{equation}
    F_{\mathrm{ad}}(\phi)
    =
    -2\pi |W|R_{\mathrm{pa}}^2(1-\cos\phi),
    \label{supp-eq:adhesion_energy}
\end{equation}
where $|W|$ is the adhesion energy per unit area, so that Eq.~\eqref{supp-eq:adhesion_energy} is simply $-|W|A_{\mathrm{bo}}$ with $A_{\mathrm{bo}}=2\pi R_{\mathrm{pa}}^2(1-\cos\phi)$ the bound area, and $\sigma$ is the effective tension of Eq.~\eqref{supp-eq:effective_tension}. The latter absorbs the $2\kappa m^2$ part of the bound-cap bending energy, which is why Eq.~\eqref{supp-eq:bound_energy} carries the linear factor $(1+2mR_{\mathrm{pa}})$ rather than the full square $(1+mR_{\mathrm{pa}})^2$. The adhesion term is the same in both geometries; the bound-cap energy is not, since a vesicle has no flat reference and a fixed rather than exchangeable area. Its counterpart is Eq.~\eqref{supp-eq:vesicle_sphere_shape}.

We have derived two solvers implementing this scheme: for a planar membrane and for a spherical vesicle. Both rest on the same variational problem for an axisymmetric membrane \cite{seifertConfigurationsFluidMembranes1997,seifertShapeTransformationsVesicles1991}, and differ in geometry, boundary conditions, and in how the bound and unbound energies of Eq.~\eqref{supp-eq:total_energy} are evaluated. We set that variational problem up once in Sec.~\ref{supp-app:general_shape_equation}. The vesicle solver of Sec.~\ref{supp-sec:vesicle_case} specialises it directly; the planar solver of Sec.~\ref{supp-sec:planar_solver} follows the equivalent formulation of \cite{agudo2021particle}, written in a different normalisation of the transverse shear.

\subsection{General axisymmetric shape equation}
\label{supp-app:general_shape_equation}


Assuming the membrane shape is axisymmetric about the $z$-axis, the unbound-membrane free energy functional
\[
F = \int_{A_{\mathrm{un}}} dA \; 2\kappa \left(M-m\right)^2 + \Sigma A_{\mathrm{un}} - \Delta P\, V_{\mathrm{un}}
\]
can be rewritten as a functional of the two profile functions $x(s)$ and $\psi(s)$ of the arc length $s$, where $x(s)$ is the radial coordinate and $\psi(s)$ the local tangent angle. Here $M$ is the mean curvature, $m$ the spontaneous curvature, and $\Sigma$ and $\Delta P$ the Lagrange multipliers conjugate to the membrane area and to the enclosed volume.

In this subsection and in the vesicle solver of Sec.~\ref{supp-sec:vesicle_case} we use the sign convention of Agudo-Canalejo and Lipowsky \cite{agudo-canalejoCriticalParticleSizes2015} for endocytic engulfment: the mother membrane carries $M>0$ (an undeformed vesicle of unit radius has $M=+1$), while the segment bound to the particle is invaginated and carries $M=-H_{\mathrm{pa}}$, with $H_{\mathrm{pa}}>0$ the particle's own mean curvature. The bending density of the bound segment is therefore $2\kappa(H_{\mathrm{pa}}+m)^2$. The planar solver of Sec.~\ref{supp-sec:planar_solver} is written with the opposite sign of $m$, i.e. with an unbound density $2\kappa(M+m)^2$; the two differ by $m\rightarrow-m$.

\begin{figure}[ht]
  \centering
  \includegraphics[width=0.8\linewidth]{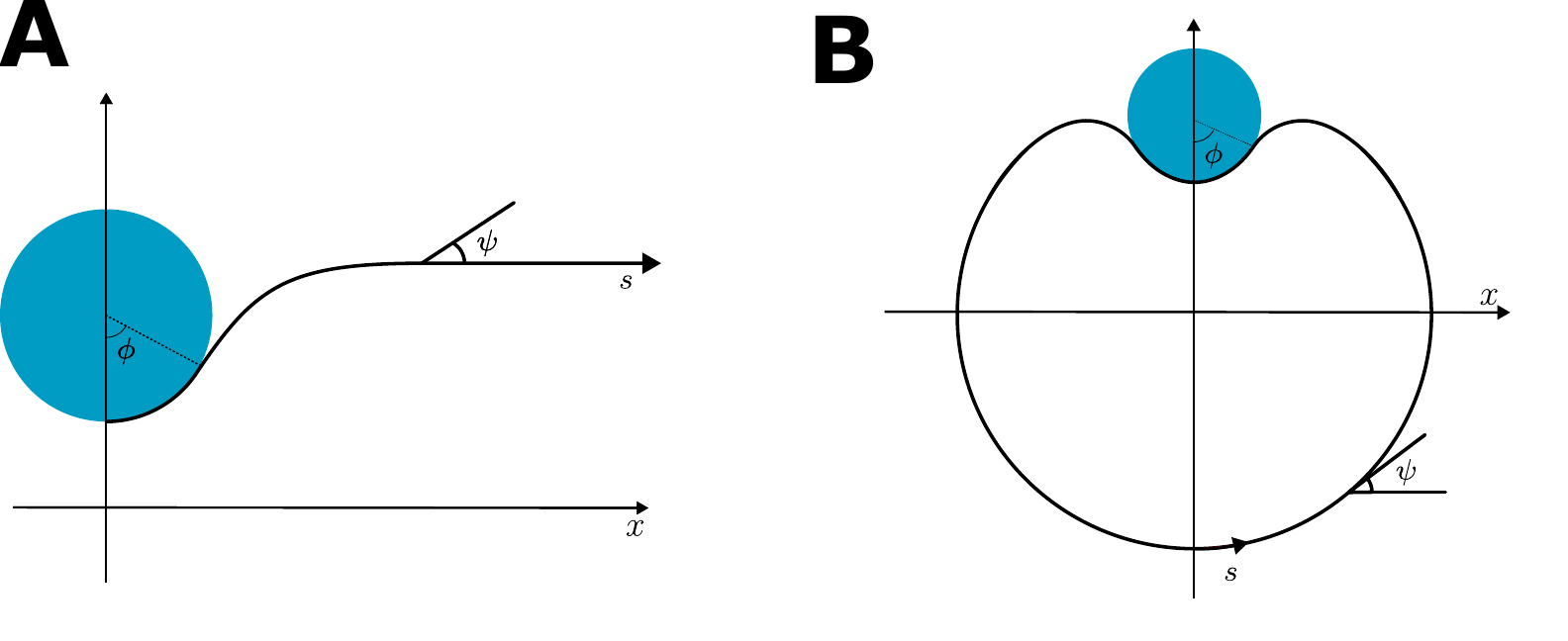}
  \caption{Geometry and parametrisation used for the axisymmetric shape equation: the arc length $s$, the radial coordinate $x(s)$, the local tangent angle $\psi(s)$ and the wrapping angle $\phi$}
  \label{supp-fig:num_schem}
\end{figure}

The mean curvature is $M=\frac12(C_1+C_2)$, with $C_1=\sin\psi/x$ the curvature transverse to the meridian and $C_2=d\psi/ds$ the curvature along it (Fig.~\ref{supp-fig:num_schem}), so that
\[
\left(M-m\right)^2 = \frac14\left(\dot\psi + \frac{\sin\psi}{x} - 2m\right)^2 .
\]
Using $dA = 2\pi x\,ds$ and $V=\pi\int x^2\sin\psi\,ds$ for the volume enclosed by the profile together with the flat disc that closes it at the contact line,
\[
F = 2\pi\kappa \int_{s_0}^{s_1} \left[\frac{x}{2}\left(\dot\psi+\frac{\sin\psi}{x}-2m\right)^2 + \frac{\Sigma}{\kappa}x - \frac{\Delta P}{2\kappa}x^2\sin\psi\right] ds.
\]
with $s_0$ and $s_1$ delimiting the unbound segment. In this form the problem is variational in the single variable $s$, which plays the role of time, and the equilibrium shape is the stationary path of the bracketed integrand. The two profile functions are not independent, however: they are tied by the geometric relation $\dot x=\cos\psi$. Imposing it with a Lagrange multiplier $\gamma(s)$ gives the Lagrangian
\[
L(s,\psi,\dot\psi,x,\dot x,\gamma) = \frac{x}{2}\left(\dot\psi+\frac{\sin\psi}{x}-2m\right)^2 + \frac{\Sigma}{\kappa}x - \frac{\Delta P}{2\kappa}x^2\sin\psi + \gamma\,\frac{\dot x - \cos\psi}{2\pi\kappa},
\]
so that $F = 2\pi\kappa\int_{s_0}^{s_1} L\,ds$. The factor $1/(2\pi\kappa)$ is a normalisation choice that gives $\gamma$ the dimensions of a force: it is the transverse shear along the profile, and we treat it as a dynamical variable in its own right, alongside $\psi$, $x$ and the meridional curvature $u=\dot\psi$.

The Euler-Lagrange equations $\frac{d}{ds}\big[\partial L/\partial\dot f_i\big]-\partial L/\partial f_i=0$ for $f_i=\psi,x$ give, after writing $u=\dot\psi$,
\begin{equation}
  \begin{cases}
    \dfrac{d\psi}{ds} = u, \\[2mm]
    \dfrac{du}{ds} = -\dfrac{u}{x}\cos\psi+\dfrac{\cos\psi\sin\psi}{x^2}+\dfrac{\gamma\sin\psi}{2\pi\kappa x}-\dfrac{\Delta P}{2\kappa}x\cos\psi, \\[2mm]
    \dfrac{d\gamma}{ds} = \pi\kappa\left(u^2-\dfrac{\sin^2\psi}{x^2}\right)+2\pi\Sigma+4\pi\kappa m\left(m-u\right)-2\pi\Delta P\,x\sin\psi, \\[2mm]
    \dfrac{dx}{ds} = \cos\psi.
  \end{cases}
  \label{supp-eq:general_shape_system}
\end{equation}
Note that $m$ does not appear explicitly in the equation for $u$: a constant shift of $M$ cancels there, and the spontaneous curvature enters the profile only through $\gamma$. The planar geometry of Sec.~\ref{supp-sec:planar_solver} imposes no volume constraint, so $\Delta P=0$ there and the corresponding terms drop out. The four variables
\begin{equation}
    \mathbf{Y}=\left(\psi,\;u,\;\gamma,\;x\right)
    \label{supp-eq:state_vector}
\end{equation}
form the state vector that is integrated. The planar solver of Sec.~\ref{supp-sec:planar_solver} integrates the same four variables and carries the height $z$ alongside them; in both geometries the remaining quantities of interest --- membrane area, enclosed volume, free energy --- are accumulated by quadrature along the profile,
\begin{equation}
    \frac{dA}{ds}=2\pi x,
    \qquad
    \frac{dV}{ds}=\pi x^2\sin\psi .
    \label{supp-eq:quadratures}
\end{equation}

The associated Hamiltonian,
\begin{equation}
\mathcal{H} = \frac{x}{2}\left[u^2-\frac{\sin^2\psi}{x^2}\right] - \left(\frac{\Sigma}{\kappa}+2m^2\right)x + 2m\sin\psi + \frac{\Delta P}{2\kappa}x^2\sin\psi + \gamma\,\frac{\cos\psi}{2\pi\kappa},
  \label{supp-eq:general_hamiltonian}
\end{equation}
is conserved along the profile since $\partial L/\partial s = 0$ (see \cite{seifertShapeTransformationsVesicles1991}, eq.~3.7). Its value is fixed by any one boundary: in the vesicle geometry the south pole forces $\mathcal{H}=0$, so that $\mathcal{H}=0$ holds along the whole profile and provides one relation between $\gamma$ and the remaining variables at the contact point. In the planar geometry the same value is fixed by the far field, where $\psi,u\rightarrow0$ and $\gamma\rightarrow2\pi\Sigma x$, so that $\mathcal{H}=0$ there as well.

\subsection{Planar solver}
\label{supp-sec:planar_solver}
We calculate the membrane shape and free-energy landscape associated with the wrapping of a particle by an initially planar membrane, following the axisymmetric formulation of
\cite{agudo2021particle}. The membrane profile is parametrised by $x(s)$ and $z(s)$, the radial and vertical coordinates, with local tangent angle $\psi$ and meridional curvature $u$. We first set up the equations for a spherical particle, and then, in Sec.~\ref{supp-sec:planar_shapes}, for the prolate spheroids and spherocylinders used in Secs.~\ref{supp-app:aspect_ratio_landscapes} and \ref{supp-app:ellipsoids}: the shape equations are the same in all three cases, and the particle enters only through the contact conditions and the bound energy.

The physical arc length of the unbound membrane is not known in advance, so we map it onto the fixed computational domain $s\in[0,1]$ through $\ell=\Omega s$, where the total arc length $\Omega$ is itself one of the unknowns; derivatives rescale as $\mathrm{d}/\mathrm{d}s=\Omega\,\mathrm{d}/\mathrm{d}\ell$. Adding the height coordinate $z$, the system reads

\begin{subequations}
\label{supp-eq:planar_shape_system_raw}
\begin{align}
\dot{\psi} &= \Omega u \\
\dot{u} &= \Omega \left( -\frac{\cos \psi}{x} u + \frac{\sin \psi \cos \psi}{x^2} + \frac{\sin \psi}{2 \cos \psi } \left[ \frac{2\sigma}{\kappa} - u^2 + \frac{\sin^2 \psi}{x^2} + \frac{4 m \sin \psi}{x} \right] \right) \\
\dot{x} &= \Omega \cos \psi \\
\dot{z} &= -\Omega \sin \psi
\end{align}
\end{subequations}

where $\sigma$ is the effective tension combining the mechanical tension $\Sigma$ and the spontaneous curvature contribution,
\begin{equation}
    \sigma=\Sigma+2\kappa m^2,
    \label{supp-eq:effective_tension}
\end{equation}
with $\kappa$ the bending rigidity and $m$ the spontaneous curvature. In our analysis we generally consider a zero-spontaneous-curvature bilayer, so $\sigma\equiv\Sigma$.


\subsubsection{Regularisation of the planar shape equations}
\label{supp-app:planar_regularisation}

The curvature equation of Eq.~\eqref{supp-eq:planar_shape_system_raw} contains the term
\begin{equation}
    \frac{\sin\psi}{2\cos\psi}
    \left(
        \frac{2\sigma}{\kappa}
        -u^2
        +\frac{\sin^2\psi}{x^2}
        +\frac{4m\sin\psi}{x}
    \right),
\end{equation}
which is numerically ill-conditioned at $\psi=\pi/2$. The singularity is only apparent, and we remove it by reintroducing the transverse shear $\gamma$ of Sec.~\ref{supp-app:general_shape_equation} through the Hamiltonian constraint
\begin{equation}
    \mathcal{H}\equiv
    \gamma\cos\psi
    -
    x\left(
        \frac{2\sigma}{\kappa}
        -u^2
        +\frac{\sin^2\psi}{x^2}
        +\frac{4m\sin\psi}{x}
    \right)
    =0.
    \label{supp-eq:app_hamiltonian}
\end{equation}
The singular term can then be replaced by
\begin{equation}
    \frac{\gamma\sin\psi}{2x},
\end{equation}
which remains finite at $\psi=\pi/2$. Differentiating
Eq.~\eqref{supp-eq:app_hamiltonian} along the membrane profile and imposing
$\mathrm{d}\mathcal{H}/\mathrm{d}\ell=0$ gives
\begin{equation}
    \frac{\mathrm{d}\gamma}{\mathrm{d}\ell}
    =
    \frac{2\sigma}{\kappa}
    +u^2
    -\frac{\sin^2\psi}{x^2}
    +4mu.
    \label{supp-eq:app_gamma}
\end{equation}
Together with the regularised curvature equation, this yields the system used
throughout, which avoids division by $\cos\psi$:
\begin{subequations}
\label{supp-eq:planar_shape_system}
\begin{align}
    \dot{\psi}
    &=
    \Omega u,
    \\
    \dot{u}
    &=
    \Omega\left(
        -\frac{\cos\psi}{x}u
        +\frac{\sin\psi\cos\psi}{x^2}
        +\frac{\gamma\sin\psi}{2x}
    \right),
    \\
    \dot{\gamma}
    &=
    \Omega\left(
        \frac{2\sigma}{\kappa}
        +u^2
        -\frac{\sin^2\psi}{x^2}
        +4mu
    \right),
    \\
    \dot{x}
    &=
    \Omega\cos\psi,
    \\
    \dot{z}
    &=
    -\Omega\sin\psi,
\end{align}
\end{subequations}
where a dot denotes differentiation with respect to $s$.

\subsubsection{Free energy accounting}

The free energy of the unbound membrane is calculated relative to a flat
membrane with the same projected area. It is accumulated alongside the
shape equations according to
\begin{equation}
    \dot{F}_{\mathrm{me}}^{\mathrm{un}}
    =
    \Omega\pi\kappa x
    \left[
        \left(
            u+\frac{\sin\psi}{x}
        \right)
        \left(
            u+\frac{\sin\psi}{x}+4m
        \right)
        +
        \frac{2\sigma}{\kappa}(1-\cos\psi)
    \right].
    \label{supp-eq:planar_unbound_energy}
\end{equation}
Integration over $s\in[0,1]$ gives
$F_{\mathrm{me}}^{\mathrm{un}}(\phi)$, completing Eq.~\eqref{supp-eq:total_energy} for the planar case.

\subsubsection{Boundary conditions}
\label{supp-sec:planar_bc}

For a prescribed wrapping angle $\phi$, the unbound membrane meets the
surface of the spherical particle at $s=0$. Smooth geometric matching gives
\begin{equation}
    \psi(0)=\phi,
    \qquad
    x(0)=R_{\mathrm{pa}}\sin\phi,
    \qquad
    z(0)=R_{\mathrm{pa}}(1-\cos\phi),
    \label{supp-eq:planar_contact_bc}
\end{equation}
where $R_{\mathrm{pa}}$ is the particle radius. The initial values
$u(0)=u_0$ and $\gamma(0)=\gamma_0$ are determined numerically (Sec.~\ref{supp-sec:planar_shooting}).

Far from the particle, the membrane approaches a flat configuration,
\begin{equation}
    \psi(\ell)\rightarrow0,
    \qquad
    u(\ell)\rightarrow0,
    \qquad
    \ell\rightarrow\infty.
    \label{supp-eq:planar_farfield}
\end{equation}
The computational domain is finite, so these conditions are imposed at its outer end as $\psi(1)=0$ and $u(1)=0$, with the domain length $\Omega$ left free for the solver to determine (Sec.~\ref{supp-sec:planar_shooting}); the converged solution places the outer boundary where the profile has already flattened. That distance is set by the membrane relaxation length
\begin{equation}
    \lambda=\sqrt{\frac{\kappa}{\sigma}} .
    \label{supp-eq:relaxation_length}
\end{equation}

\paragraph{Decay of the far field}

Far from the particle, $\psi$ and $u$ are small and $x$ is large, so in the raw curvature equation (Eq.~\eqref{supp-eq:planar_shape_system_raw}) the curvature terms $\sin\psi\cos\psi/x^2$ and $\sin^2\psi/x^2$ are negligible next to the terms linear in $\psi$. With $m=0$ and expanding $\cos\psi\approx1$, $\sin\psi/(2\cos\psi)\approx\psi/2$ to linear order,
\begin{equation}
    \frac{\mathrm{d}u}{\mathrm{d}\ell} \approx \frac{\sigma}{\kappa}\,\psi,
    \qquad
    \frac{\mathrm{d}\psi}{\mathrm{d}\ell} = u,
\end{equation}
so that $\psi$ obeys $\mathrm{d}^2\psi/\mathrm{d}\ell^2 = (\sigma/\kappa)\,\psi$, whose bounded solution is the decaying branch $\psi(\ell)\propto e^{-\ell/\lambda}$, the standard relaxation of a tense membrane. The deformation induced by the particle therefore relaxes over a length of order $\lambda$, which is what makes it legitimate to impose Eq.~\eqref{supp-eq:planar_farfield} at a finite $\Omega$ of a few $\lambda$ rather than asymptotically.

\subsubsection{Non-spherical particles}
\label{supp-sec:planar_shapes}

The shape equations, Eq.~\eqref{supp-eq:planar_shape_system}, make no reference to the particle: exactly as in the vesicle solver, the particle enters only through its contact state and its bound energy. Writing $\theta\in[0,\pi]$ for the wrapping coordinate, which measures how far the contact line has advanced along the particle meridian and reduces to the wrapping angle $\phi$ for a sphere, the contact conditions of Eq.~\eqref{supp-eq:planar_contact_bc} generalise to
\begin{equation}
    \psi(0)=\psi_0(\theta),
    \qquad
    x(0)=x_0(\theta),
    \qquad
    z(0)=z_0(\theta),
    \label{supp-eq:planar_contact_bc_general}
\end{equation}
with $\psi_0$ the local tangent angle of the particle surface at the contact line, and the bound and adhesion energies of Eqs.~\eqref{supp-eq:bound_energy}--\eqref{supp-eq:adhesion_energy} to
\begin{equation}
    F_{\mathrm{me}}^{\mathrm{bo}}(\theta)
    =
    \int_{A_{\mathrm{bo}}} 2\kappa H_{\mathrm{pa}}\left(H_{\mathrm{pa}}+2m\right)\,dA
    +
    \sigma\left(A_{\mathrm{bo}}-\pi x_0^2\right),
    \qquad
    F_{\mathrm{ad}}(\theta)=-|W|\,A_{\mathrm{bo}}(\theta),
    \label{supp-eq:planar_bound_energy_general}
\end{equation}
where $H_{\mathrm{pa}}$ is the particle's own mean curvature and $\pi x_0^2$ is the area that the bound cap projects onto the far-field plane. The constant $2\kappa m^2$ is omitted here, as it is in Eq.~\eqref{supp-eq:planar_unbound_energy}, being absorbed into the effective tension of Eq.~\eqref{supp-eq:effective_tension}. For a sphere, $\psi_0=\theta=\phi$ and $H_{\mathrm{pa}}=1/R_{\mathrm{pa}}$, and Eqs.~\eqref{supp-eq:planar_contact_bc_general}--\eqref{supp-eq:planar_bound_energy_general} reduce to Eqs.~\eqref{supp-eq:planar_contact_bc} and \eqref{supp-eq:bound_energy}--\eqref{supp-eq:adhesion_energy}.

\paragraph{Prolate spheroid} With equatorial semi-axis $a$ and polar semi-axis $c$, and writing $g(\theta)=a^2\cos^2\theta+c^2\sin^2\theta$,
\begin{equation}
    x_0=a\sin\theta,
    \quad
    z_0=c\left(1-\cos\theta\right),
    \quad
    \psi_0=\operatorname{atan2}\left(c\sin\theta,\;a\cos\theta\right),
    \quad
    A_{\mathrm{bo}}=2\pi\!\int_0^{\theta}\! a\sin t\,\sqrt{g(t)}\,dt,
\end{equation}
with the particle mean curvature
\begin{equation}
    H_{\mathrm{pa}}(\theta)=\frac12\left[\frac{ac}{g^{3/2}}+\frac{c}{a\sqrt{g}}\right] .
\end{equation}
The aspect ratio is $c/a$, and $a=c$ recovers the sphere.

\paragraph{Spherocylinder} A rounded rod is described by its cap radius $a$ and the length $h$ of its cylindrical section, so that its total length is $h+2a$ and its diameter $2a$. The wrapping coordinate is mapped linearly onto the meridional arc length,
\begin{equation}
    \varsigma(\theta)=\frac{\theta}{\pi}\left(\pi a+h\right),
\end{equation}
and the value of $\varsigma$ selects the piece of surface on which the contact line sits: the lower cap for $\varsigma\leq\pi a/2$, the cylindrical wall for $\pi a/2<\varsigma\leq\pi a/2+h$, and the upper cap beyond. On the caps $H_{\mathrm{pa}}=1/a$ and on the wall $H_{\mathrm{pa}}=1/2a$, and $x_0$, $z_0$, $\psi_0$ and $A_{\mathrm{bo}}$ follow piecewise from the same construction; on the wall in particular $\psi_0=\pi/2$ and $x_0=a$, so the contact line advances at constant radius. We characterise the rod by the aspect ratio
\begin{equation}
    AR=\frac{h+2a}{2a},
    \label{supp-eq:rod_aspect_ratio}
\end{equation}
the ratio of total length to diameter, so that $AR=1$ is a sphere. The families compared in Sec.~\ref{supp-app:aspect_ratio_landscapes} are built at fixed particle area $A=4\pi a^2+2\pi a h$: eliminating $h$ gives $A=4\pi a^2 AR$, so that
\begin{equation}
    a=\sqrt{\frac{A}{4\pi\,AR}},
    \qquad
    h=2a\left(AR-1\right)
    \label{supp-eq:rod_equal_area}
\end{equation}
generates the whole equal-area family from a single target area.

\subsection{Vesicle solver}\label{supp-sec:vesicle_case}

We now detail the formulation for a closed lipid vesicle engulfing a rigid particle. Unlike the planar membrane, which acts as an infinite lipid reservoir held at a fixed imposed tension, a vesicle is a finite system: its total area $A=4\pi R_{\mathrm{ves}}^2$ is fixed, and the tension is no longer a control parameter but a Lagrange multiplier determined by that constraint. The enclosed volume may be either left free or constrained in the same way, with $\Delta P$ the conjugate multiplier. The outer boundary condition at infinity of the planar case is replaced by closure at the opposite pole of the vesicle.

For a fixed wrapping angle $\phi$ the equilibrium profile of the unbound membrane extremizes
\begin{equation}
    F = E_{\mathrm{un}} + \Sigma\left(A_{\mathrm{un}}-A^\star\right) - \Delta P\left(V_{\mathrm{un}}-V^\star\right),
    \label{supp-eq:vesicle_functional}
\end{equation}
where $E_{\mathrm{un}}\equiv F_{\mathrm{me}}^{\mathrm{un}}$ is the bending energy of the unbound segment and $A^\star$, $V^\star$ are the prescribed area and volume. Up to the constant $\Sigma A^\star-\Delta P\,V^\star$, which does not affect the shape equations, this is the functional of Sec.~\ref{supp-app:general_shape_equation}, so the profile obeys the shape system of Eq.~\eqref{supp-eq:general_shape_system}. Note that the multiplier terms belong to the functional being extremized, not to its value: at the solution the constraints are satisfied and $F=E_{\mathrm{un}}$, so the energy landscape of Sec.~\ref{supp-sec:classifying_states} must be built from the bending energy alone (Sec.~\ref{supp-sec:vesicle_energy}).

\subsubsection{Dimensionless variables and rescaled domain}
\label{supp-sec:vesicle_units}

All lengths are measured in units of the vesicle radius $R_{\mathrm{ves}}$ and all energies in units of the bending rigidity, so that $\kappa=1$ and the total vesicle area is $4\pi$. In particular
\begin{equation}
    r_{\mathrm{pa}}=\frac{R_{\mathrm{pa}}}{R_{\mathrm{ves}}},
    \quad
    \sigma=\frac{\Sigma R_{\mathrm{ves}}^2}{\kappa},
    \quad
    p=\frac{\Delta P R_{\mathrm{ves}}^3}{\kappa},
    \quad
    w=\frac{|W| R_{\mathrm{ves}}^2}{\kappa},
    \label{supp-eq:vesicle_units}
\end{equation}
and the spontaneous curvature is given in units of $1/R_{\mathrm{ves}}$. As in the planar case, the unknown physical arc length $\ell$ is mapped onto the fixed computational domain $s\in[0,1]$ through $\ell=\Omega s$, where the total contour length $\Omega$ is itself one of the unknowns. Here it is a genuine physical length rather than the position of an outer boundary, since the profile closes on the vesicle. The system integrated is then
\begin{subequations}
\label{supp-eq:vesicle_shape_system}
\begin{align}
    \dot\psi &= \Omega u, \\
    \dot u &= \Omega\left(-\frac{\cos\psi}{x}u+\frac{\sin\psi\cos\psi}{x^2}+\frac{\gamma\sin\psi}{2\pi x}-\frac{p}{2}x\cos\psi\right), \\
    \dot\gamma &= \Omega\left[\pi\left(u^2-\frac{\sin^2\psi}{x^2}\right)+2\pi\left(\sigma+2m^2-2mu\right)-2\pi p\,x\sin\psi\right], \\
    \dot x &= \Omega\cos\psi,
\end{align}
\end{subequations}
with the area and volume accumulated alongside it as $\dot A=2\pi\Omega x$ and $\dot V=\pi\Omega x^2\sin\psi$. The conserved Hamiltonian of Eq.~\eqref{supp-eq:general_hamiltonian} reads, in these units,
\begin{equation}
    \mathcal{H}=\frac{x}{2}\left(u^2-\frac{\sin^2\psi}{x^2}\right)
    -\left(\sigma+2m^2\right)x
    +2m\sin\psi
    +\frac{p}{2}x^2\sin\psi
    +\frac{\gamma\cos\psi}{2\pi}
    =0 .
    \label{supp-eq:vesicle_hamiltonian}
\end{equation}

\subsubsection{Particle shapes}
\label{supp-sec:vesicle_shapes}

The shape equations themselves are independent of the particle: a particle of a given shape enters the problem only through three functions of the wrapping angle $\phi$, namely the contact state $\left(x_0,z_0,\psi_0\right)$ on its surface, with $\psi_0$ the local tangent angle of the particle surface at the contact line, the bound area $A_{\mathrm{bo}}(\phi)$, and the bound bending energy $E_{\mathrm{bo}}^{\mathrm{be}}(\phi)$. The solver is therefore shape-agnostic; the vesicle results reported here are for spheres, but the same machinery covers prolate spheroids and spherocylinders. The non-spherical results of Secs.~\ref{supp-app:aspect_ratio_landscapes} and \ref{supp-app:ellipsoids} were obtained with the planar solver of Sec.~\ref{supp-sec:planar_solver}, not with this one.

For a \emph{sphere} of radius $r_{\mathrm{pa}}$,
\begin{equation}
    \psi_0=\phi,
    \quad
    x_0=r_{\mathrm{pa}}\sin\phi,
    \quad
    A_{\mathrm{bo}}=2\pi r_{\mathrm{pa}}^2\left(1-\cos\phi\right),
    \quad
    E_{\mathrm{bo}}^{\mathrm{be}}=4\pi\left(1+m\,r_{\mathrm{pa}}\right)^2\left(1-\cos\phi\right),
    \label{supp-eq:vesicle_sphere_shape}
\end{equation}
the last expression being eq.~S4 of Agudo-Canalejo and Lipowsky \cite{agudo-canalejoCriticalParticleSizes2015}. For a \emph{spheroid} with equatorial semi-axis $a$ and polar semi-axis $c$, writing $g(\phi)=a^2\cos^2\phi+c^2\sin^2\phi$ and $\operatorname{atan2}$ for the two-argument arctangent, which keeps $\psi_0$ on the correct branch for $\phi>\pi/2$,
\begin{equation}
    x_0=a\sin\phi,
    \quad
    \psi_0=\operatorname{atan2}\left(c\sin\phi,\;a\cos\phi\right),
    \quad
    A_{\mathrm{bo}}=2\pi\!\int_0^{\phi}\! a\sin t\,\sqrt{g(t)}\,dt,
\end{equation}
and, with the particle mean curvature $H_{\mathrm{pa}}=\frac12\left[ac\,g^{-3/2}+c\,a^{-1}g^{-1/2}\right]$,
\begin{equation}
    E_{\mathrm{bo}}^{\mathrm{be}}=4\pi\!\int_0^{\phi}\!\left(H_{\mathrm{pa}}(t)+m\right)^2 a\sin t\,\sqrt{g(t)}\,dt ;
\end{equation}
$a=c$ recovers Eq.~\eqref{supp-eq:vesicle_sphere_shape} exactly. For a \emph{spherocylinder} of cap radius $a$ and half-height $c\geq a$, the wrapping angle is mapped linearly onto the meridional arc length, $\varsigma=\left(\phi/\pi\right)\left[\pi a+2(c-a)\right]$, so that $\phi=\pi$ again corresponds to complete wrapping. The value of $\varsigma$ selects the piece of the surface on which the contact line sits --- the lower cap for $\varsigma\leq\pi a/2$, the cylindrical wall for $\pi a/2<\varsigma\leq\pi a/2+2(c-a)$, and the upper cap beyond --- and $A_{\mathrm{bo}}$ and $E_{\mathrm{bo}}^{\mathrm{be}}$ follow by integrating over the wrapped portion, the caps and the wall contributing bending densities $2\left(1/a+m\right)^2$ and $2\left(1/2a+m\right)^2$ respectively.

\subsubsection{Boundary conditions}
\label{supp-sec:vesicle_bc}

The integration domain runs from the south pole of the vesicle to the contact line, which is placed at $s=1$ by construction. At the south pole, symmetry gives
\begin{equation}
    \psi=0,
    \qquad
    x=0,
    \qquad
    u=u_0 ,
\end{equation}
with $u_0$ unknown. Evaluating Eq.~\eqref{supp-eq:vesicle_hamiltonian} there, the two curvature terms cancel in the limit $x\rightarrow0$, since $\sin^2\psi/2x\rightarrow u_0^2x/2$, and every remaining term except the last vanishes with $x$; what survives is $\gamma\cos\psi=0$, so $\gamma=0$ at the pole and, by conservation, $\mathcal{H}=0$ along the entire profile.

At the contact line the unbound profile must join the bound segment, which imposes exactly three conditions,
\begin{equation}
    \psi^\star=\pi+\psi_0(\phi),
    \qquad
    x^\star=x_0(\phi),
    \qquad
    A^\star=4\pi-A_{\mathrm{bo}}(\phi),
    \label{supp-eq:vesicle_contact_bc}
\end{equation}
the first two being the geometric matching to the particle surface and the third the fixed-area constraint. The meridional curvature $u^\star$ at the contact line is \emph{not} prescribed: it, and the transverse shear $\gamma^\star$, are free unknowns determined by the solve (Sec.~\ref{supp-sec:gamma_unknown}).

\subsubsection{Expansion at the south pole}
\label{supp-sec:vesicle_pole}

At the south pole the radial coordinate $x$ vanishes and the $1/x$ and $1/x^2$ terms of Eq.~\eqref{supp-eq:vesicle_shape_system} are singular, so the integration cannot be started there. We instead expand the solution in the physical arc length $\ell=\Omega s$ about $\ell=0$ and use the series to step to a small but finite radius, in practice the value of $\ell$ at which $x\simeq3.5\times10^{-2}$. Substituting the expansions into Eq.~\eqref{supp-eq:vesicle_shape_system} and matching order by order gives
\begin{subequations}
\label{supp-eq:vesicle_series}
\begin{align}
    x(\ell) &= \ell-\frac{u_0^2}{6}\ell^3, \\
    \psi(\ell) &= u_0\ell+\frac{b_3}{6}\ell^3, \\
    u(\ell) &= u_0+\frac{b_3}{2}\ell^2, \\
    \gamma(\ell) &= 2\pi\left(g_1\ell+\frac{g_3}{6}\ell^3\right),
\end{align}
\end{subequations}
with the coefficients
\begin{equation}
    g_1=\sigma+2m^2-2mu_0,
    \qquad
    b_3=\frac{3}{4}\left(u_0g_1-\frac{p}{2}\right),
    \qquad
    g_3=2b_3\left(\frac{u_0}{3}-m\right)-2pu_0 .
    \label{supp-eq:vesicle_series_coeff}
\end{equation}
For a curvature-symmetric bilayer at free volume ($m=p=0$) these reduce to $g_1=\sigma$, $b_3=\tfrac34\sigma u_0$ and $g_3=\tfrac12\sigma u_0^2$. The coefficients of Eq.~\eqref{supp-eq:vesicle_series_coeff} are the ones for which the Hamiltonian of Eq.~\eqref{supp-eq:vesicle_hamiltonian} vanishes identically through $\mathcal{O}(\ell^3)$; this is a useful check, since a series that is correct only to lower order leaves a residual in $\mathcal{H}$ that the integrator then propagates along the whole profile.

\subsubsection{Free energy accounting}
\label{supp-sec:vesicle_energy}

As noted below Eq.~\eqref{supp-eq:vesicle_functional}, the multiplier terms vanish at the solution and must not be carried into the energy landscape. The energy of the unbound segment is therefore its bending energy alone,
\begin{equation}
    E_{\mathrm{un}}=2\pi\Omega\int_0^1 \frac{x}{2}\left(u+\frac{\sin\psi}{x}-2m\right)^2 ds ,
\end{equation}
and the total free energy of the wrapping process at adhesion strength $w$ is
\begin{equation}
    F_{\mathrm{tot}}(\phi)=E_{\mathrm{un}}(\phi)+E_{\mathrm{bo}}^{\mathrm{be}}(\phi)-w\,A_{\mathrm{bo}}(\phi),
    \label{supp-eq:vesicle_energy}
\end{equation}
which is Eq.~\eqref{supp-eq:total_energy} for the vesicle case, with $F_{\mathrm{me}}^{\mathrm{un}}=E_{\mathrm{un}}$, $F_{\mathrm{me}}^{\mathrm{bo}}=E_{\mathrm{bo}}^{\mathrm{be}}$ and $F_{\mathrm{ad}}=-w\,A_{\mathrm{bo}}$, the bound-cap and adhesion terms of Sec.~\ref{supp-sec:vesicle_shapes} replacing the planar Eqs.~\eqref{supp-eq:bound_energy}--\eqref{supp-eq:adhesion_energy}. In particular there is no tension contribution: in the vesicle the area is fixed rather than exchanged with a reservoir, and $\sigma$ is an output of the solve rather than an input. Because $w$ enters Eq.~\eqref{supp-eq:vesicle_energy} only through the last term, one sweep in $\phi$ yields the landscape for every adhesion strength: it is enough to store $E_{\mathrm{un}}$, $E_{\mathrm{bo}}^{\mathrm{be}}$ and $A_{\mathrm{bo}}$ and to recombine them in post-processing.

\subsection{Solving the boundary-value problem}
\label{supp-app:multi_shooting}
\label{supp-sec:planar_shooting}

Both geometries lead to the same kind of numerical problem. The shape equations are known at the two ends of the profile but not in between: the contact conditions fix $\psi$ and $x$ (and $z$, in the planar case) where the membrane meets the particle, the outer end is fixed by the far field or by the pole, and the initial curvature, the transverse shear and the domain length are unknown. This subsection describes the machinery common to both solvers; the geometry-specific ingredients are those of Secs.~\ref{supp-sec:planar_solver} and \ref{supp-sec:vesicle_case}.

\subsubsection{The transverse shear is an unknown, not a boundary condition}
\label{supp-sec:gamma_unknown}

At the contact line the Hamiltonian constraint $\mathcal{H}=0$ is one relation among the unknowns, and it can be solved for the transverse shear. In the planar case,
\begin{equation}
    \gamma_0 = \frac{x(0)}{\cos\psi_0}\left[\frac{2\sigma}{\kappa}-u_0^2+\frac{\sin^2\psi_0}{x(0)^2}+\frac{4m\sin\psi_0}{x(0)}\right], \qquad \cos\psi_0\neq0 ,
    \label{supp-eq:gamma0_of_u0}
\end{equation}
and in the vesicle case, for a sphere,
\begin{equation}
    \gamma^\star=-2\pi r_{\mathrm{pa}}\,\sigma\tan\phi .
    \label{supp-eq:vesicle_gamma_closed}
\end{equation}
Both carry a prefactor $1/\cos\psi_0$ that is singular wherever the particle surface is tangent to the symmetry axis: a sphere at $\phi=\pi/2$, the equator of a spheroid, and the entire cylindrical wall of a spherocylinder, where $\psi_0=\pi/2$ over a finite range of the wrapping coordinate. Near such a point the closed form changes sign through a pole, and a solver seeded from it is pushed onto a different branch.

We therefore never impose these expressions as boundary conditions. The shear at the contact line is carried as an explicit unknown and $\mathcal{H}=0$ is added to the residuals instead, which stays smooth through $\cos\psi_0=0$: there the $\gamma\cos\psi$ term drops out and the same equation constrains the remaining unknowns. Equations~\eqref{supp-eq:gamma0_of_u0} and \eqref{supp-eq:vesicle_gamma_closed} are used only to seed cold starts, in the vesicle case with $1/\cos\psi^\star$ replaced by the regularised $\cos\psi^\star/(\cos^2\psi^\star+\epsilon^2)$; warm starts inherit the converged shear of the neighbouring solve. This is what makes the perpendicular contact angle, the equator of a spheroid and the cylindrical wall of a rod numerically accessible.

\subsubsection{Multiple shooting}

Single shooting integrates the shape equations across the entire domain in one pass. Small errors in the initial data then grow exponentially along the profile, so we divide $[0,1]$ into $N$ sub-intervals separated by $N-1$ interior nodes, integrate each independently, and require continuity of the state vector at every node,
\begin{equation}
    \mathbf{R}_i=\mathbf{Y}_{\mathrm{left}}(s_i)-\mathbf{Y}_{\mathrm{right}}(s_i)=0 .
\end{equation}
In the planar geometry all $N$ sub-intervals are integrated forward from the contact line. In the vesicle geometry the shooting is two-sided: the first $N-1$ run forward from the south-pole series of Eq.~\eqref{supp-eq:vesicle_series}, the last runs backward from the contact line at $s=1$, whose state is $(\psi^\star,u^\star,\gamma^\star,x^\star)$, and the two branches meet at the last interior node.

\paragraph{Unknowns and residuals} The two parameter vectors are

\begin{center}
\begin{tabular}{lll}
\hline
 & planar (Sec.~\ref{supp-sec:planar_solver}) & vesicle (Sec.~\ref{supp-sec:vesicle_case}) \\
\hline
state vector & $(\psi,u,\gamma,x,z)$ & $(\psi,u,\gamma,x)$ \\
scalar unknowns & $\Omega,\;u_0,\;\gamma_0$ & $\Omega,\;\sigma,\;u_0,\;u^\star,\;\gamma^\star,\;p$ \\
node unknowns & $5(N-1)$ & $4(N-1)$ \\
continuity residuals & $5(N-1)$ & $4(N-1)$ \\
matching residuals & $\psi(1)=0$, $u(1)=0$ & $\psi$, $u$, $x$ at the junction \\
constraint residuals & $\mathcal{H}=0$ at the contact & $\mathcal{H}(1)=0$; area; volume \\
\hline
\end{tabular}
\end{center}

\noindent
so that both systems are square: $5N-2$ equations in the planar case and $4N+2$ in the vesicle case. The area residual is $A(1)=A^\star$, evaluated on the quadrature of Eq.~\eqref{supp-eq:quadratures}, which starts at the series radius and therefore omits the small polar cap below it; the volume residual is either $p=0$, when the volume is left free, or $V(1)-V_{\mathrm{bo}}=v\,\tfrac43\pi$, when a reduced volume $v$ is prescribed. Here $V(1)$ is the auxiliary volume enclosed by the profile and by the disc that closes it at the contact line, and $V_{\mathrm{bo}}$ the volume of the particle cap adjacent to the bound segment, so that $V(1)-V_{\mathrm{bo}}$ is the vesicle volume and $v$ its value relative to a sphere of the same area.

Both systems are solved with a bounded trust-region least-squares method. Trajectories that leave the physical region are terminated by guards --- $x$ below a floor, any state component above a ceiling, the domain length above its bound --- and reported to the optimizer as a large residual, which keeps it inside the feasible box.

\subsubsection{Continuation}
\label{supp-sec:continuation}

We need the solution over a whole range of wrapping angles, and the initial guess matters: the solver converges only from close enough to the root, and the landscapes are genuinely bistable in places (Sec.~\ref{supp-sec:classifying_states}), so a poor guess can land on the wrong branch or fail outright. We therefore sweep the angles $\{\phi_i\}$ by continuation:
\begin{enumerate}
    \item Start at a $\phi_0$ where a good initial guess is available, for instance a small angle, where the contact patch is small and the membrane is only weakly perturbed.
    \item Solve there, then step to $\phi_1=\phi_0+\Delta\phi$ using the converged parameter vector as the initial guess.
    \item If the solver converges, keep the result and step again; if not, halve $\Delta\phi$ and retry from the last converged angle. A step that converges in few iterations allows $\Delta\phi$ to be grown again.
    \item Repeat until the desired range of $\phi$ is covered.
\end{enumerate}
Seeding each solve with the previous converged one, rather than with a fixed guess, keeps every step inside the basin of convergence as long as $\Delta\phi$ is small enough. The same device carries the solvers along the other parameters: an elongated particle is reached by ramping its aspect ratio from the sphere one warm-started step at a time, and a prescribed reduced volume by first solving the free-volume problem and then walking $v$ towards the target, since $p$ has to grow from zero while the shape rearranges.

\paragraph{Practical tips}
\begin{itemize}
    \item Check that the converged domain length is comfortably larger than $\lambda$ and does not sit against its upper bound: if it does, the far-field conditions are being imposed while the profile is still bending, and the bound should be raised.
    \item Monitor $\mathcal{H}(s)$ along the integrated trajectory: since it is conserved (Sec.~\ref{supp-app:general_shape_equation}), it should stay close to zero throughout if the integration is accurate.
    \item Repeatedly having to shrink $\Delta\phi$ can signal a fold in $u_0(\phi)$ rather than a numerical problem; check the energy landscape for bistability (Sec.~\ref{supp-sec:classifying_states}) before treating it as one.
\end{itemize}

\subsection{Classifying wrapping states}
\label{supp-sec:classifying_states}

Once the total free energy landscape $F_{\mathrm{tot}}(\phi)$ is computed, the equilibrium state follows from the arrangement of its minima. We use the four states of Agudo-Canalejo and Lipowsky \cite{agudo-canalejoCriticalParticleSizes2015}:
\begin{enumerate}
    \item \textbf{Free state} ($F_{\mathrm{st}}$): the energy increases monotonically with the wrapping angle, the only minimum is the unbound state at $\phi=0$, and the membrane rejects the particle.
    \item \textbf{Partially wrapped state} ($P_{\mathrm{st}}$): the single minimum lies at an intermediate angle $0<\phi<\pi$, where the adhesion gain is balanced by the bending and tension cost of wrapping further, and the particle is held partially wrapped.
    \item \textbf{Completely engulfed state} ($C_{\mathrm{st}}$): the energy decreases monotonically, the minimum sits at $\phi=\pi$, and the particle is fully engulfed without a kinetic barrier.
    \item \textbf{Bistable} ($B_{\mathrm{st}}$): two minima coexist, separated by a barrier. Which one is realised depends on the barrier height and on how the system was prepared; a barrier comparable to the thermal energy $k_{\mathrm{B}}T$ can be crossed by fluctuations, whereas a barrier of many $k_{\mathrm{B}}T$ traps the particle in whichever minimum it reaches first.
\end{enumerate}
The second minimum of a bistable landscape may be either partially wrapped, as in the planar sweeps of the main text, or completely engulfed, as in the vesicle state diagram of Fig.~\ref{supp-fig:vesicle_validation}; both are labelled $B_{\mathrm{st}}$.

The vesicle state diagram uses all four labels. The planar sweeps reported here resolve only three of them --- free, bistable and completely engulfed --- because a stable partially wrapped minimum is detected as one of the two wells of the bistable class rather than as a class of its own; it appears explicitly as the ``partially wrapped'' region of the reference diagrams reproduced in Fig.~\ref{supp-fig:rpa_plot_comparison}(A, B). Which of the four states occurs depends on tension, particle size and particle shape, analysed in the respective sections of the main text.

\section{FreeDTS}\label{supp-app:freedts}
We study the wrapping of rigid and deformable particles by a fluid membrane using the dynamically triangulated surface framework as implemented in FreeDTS\cite{pezeshkianMesoscaleSimulationBiomembranes2024}. The membrane is represented by vertices connected into a triangular mesh, and its configurations are sampled using Monte Carlo moves. Vertex displacements change the shape, while edge flips change the mesh connectivity, allowing the membrane to behave as a fluid surface. The membrane forms a periodic patch in the $xy$ plane. Edge lengths are restricted to $[\ell_{0},\sqrt{3}\ell_{0}]$, and moves that violate the mesh distance or face-angle constraints are rejected.

\subsection{Particle geometry}
We consider rigid sphere, rigid spherocylinder, and deformable particle. The rigid particle is represented by analytical geometries, while each deformable particle is represented by a separate triangulated surface.

\paragraph{Spherical particle}
A spherical particle is defined by its centre $\mathbf r_c$ and effective radius $R$. For a membrane vertex at $\mathbf x_i$, the distance to the particle centre is
\begin{equation}
    r_i = |\mathbf x_i-\mathbf r_c|,
\end{equation}
The particle can translate, while its radius remains fixed. Its orientation does not affect the interaction because of spherical symmetry.

\paragraph{Spherocylinder particle}
The rod is represented as a spherocylinder with centre $\mathbf r_c$, unit axis vector $\hat{\mathbf u}$, effective radius $r_{rod}$, and cylindrical length $L$.
Its end-to-end length is $h=L + 2r_{rod}$. For a membrane vertex at $\mathbf x_i$, the closest point on the finite axis segment is
\begin{equation}
    \mathbf q_i = \mathbf r_c+t_i\hat{\mathbf u},
    \qquad
    t_i = \max\!\left[
        -\frac{L}{2},
        \min\!\left(
            \frac{L}{2},
            (\mathbf x_i-\mathbf r_c)\cdot\hat{\mathbf u}
        \right)
    \right].
\end{equation}
The distance to this segment is
\begin{equation}
    r_i = |\mathbf x_i-\mathbf q_i|,
\end{equation}
 The surface $r_i=r_{rod}$ therefore forms a cylinder with hemispherical caps. In the simulation, the rod can translate and rotate about its centre. For both rigid particle geometries, vertex positions are evaluated using periodic images relative to the particle centre in the $xy$ plane.

\paragraph{Soft particle}

A soft particle is represented by a closed triangular mesh, separate from the wrapping membrane. Its shape can change through displacements of its vertices during wrapping. Edge flips are disabled on the particle surface but remain allowed on the wrapping membrane\cite{kantor1986statistical}. The particle therefore retains its initial network connectivity and cannot relax in-plane deformation through neighbour exchanges. This gives a deformable shell with fixed connectivity, following the tethered-surface description.

The membrane-particle interaction is evaluated between vertices belonging to the two surfaces. If $\mathbf x_i$ is a vertex of the wrapping membrane and $\mathbf y_j$ is a vertex of the particle, their separation is
\begin{equation}
    r_{ij}=|\mathbf x_i-\mathbf y_j|,
\end{equation}
The energy terms governing particle deformation and membrane-particle adhesion are described below.

\subsection{Energy}

The total free energy of the planar membrane is
\begin{equation}
    E_{\mathrm{M}} = E_{\mathrm{bending}} + E_{\mathrm{tension}} + E_{\mathrm{adhesion}},
\end{equation}
The first three terms describe membrane bending, the imposed frame tension, and membrane-particle adhesion, respectively. The particle contribution is constant and omitted for the rigid particle. For a soft
particle, it includes its own bending energy.

\paragraph{Bending energy}

The membrane bending energy is given by the discretized Helfrich Hamiltonian,
\begin{equation}
    E_{\mathrm{bending}}
    = \frac{\kappa}{2}\sum_i A_i(2H_i-C_0)^2,
    \label{supp-eq:freedts_bending}
\end{equation}
where $\kappa$ is the bending rigidity, $H_i=(c_1^{(i)}+c_2^{(i)})/2$ is the local mean curvature, and $C_0$ is the spontaneous curvature parameter. The principal curvatures are calculated using the discrete shape operator. The area assigned to a vertex is one third of the total area of its adjacent triangles, $A_i=\tfrac13\sum_{t\ni i}A_t$. We set $C_0=0$ and the Gaussian curvature modulus to zero, so that $E_{\mathrm{bending}}=2\kappa\sum_i A_iH_i^2$.

\paragraph{Frame tension}

We impose a frame tension $\sigma$ using box moves. Each trial move scales
$L_x$, $L_y$, and the membrane coordinates in the $xy$ plane by the same factor, while keeping $L_z$ fixed. The tension contribution to the energy is
\begin{equation}
    E_{\mathrm{tension}} = -\,\sigma\, A_{\mathrm{p}}, \qquad A_{\mathrm{p}} = L_x L_y,
\end{equation}
Here, $A_{\mathrm p}=L_xL_y$ is the projected area of the simulation box. A positive $\sigma$ favours a larger projected area.

\paragraph{Adhesive energy}
\paragraph{Membrane-particle interaction}
For a rigid sphere or rod, let $r_i$ be the distance from membrane
vertex $i$ to the particle centre or the closest point on the rod-axis segment, respectively. We use a Lennard-Jones potential
\begin{equation}
v(r)=w\left[
    \left(\frac{r_0}{r}\right)^{12}
    -2\left(\frac{r_0}{r}\right)^6
\right],
\end{equation}
where $w$ is the interaction-strength parameter used in the figures
and $r_0$ is the position of the potential minimum. We set $r_0=R$
for a sphere and $r_0=c$ for a rod.

On the adhesive side of the membrane, the potential is truncated and shifted at $r_{\mathrm c}=\lambda_{\mathrm c}r_0$, where
$\lambda_{\mathrm c}=1.5$ is the cutoff factor:
\begin{equation}
u_{\mathrm{att}}(r)=
\begin{cases}
v(r)-v(r_{\mathrm c}), & r\leq r_{\mathrm c},\\
0, & r>r_{\mathrm c}.
\end{cases}
\end{equation}
The shift makes the potential continuous at the cutoff. Its minimum
remains at $r=r_0$, with value $-w-v(r_{\mathrm c})$.

On the nonadhesive side of the membrane, vertices interact with
the particle only through the repulsive core:
\begin{equation}
u_{\mathrm{rep}}(r)=
\begin{cases}
v(r)+w, & r<r_0,\\
0, & r\geq r_0.
\end{cases}
\end{equation}

The total membrane-particle interaction energy is
\begin{equation}
E_{\mathrm{adhesion}}
=\sum_i\left[
s_i u_{\mathrm{att}}(r_i)
+(1-s_i)u_{\mathrm{rep}}(r_i)
\right].
\end{equation}
The membrane normals are oriented towards the side on which the
particle is initially placed. We set $s_i=1$ when the local normal
points towards the particle centre or the closest point on the
rod-axis segment, and $s_i=0$ otherwise. The interaction energy is
summed over membrane vertices without area weighting, so $w$ has
units of energy.

\subsection{Monte Carlo scheme}

We sample membrane and particle configurations using Monte Carlo
moves. Vertex displacements change surface shape, while edge flips
rearrange the connectivity of the wrapping membrane. The rigid sphere undergoes trial translations. The freely rotating rod additionally undergoes orientation changes. Soft particle deforms through vertex displacements, while its connectivity remains fixed. In-plane box moves allow the projected area to fluctuate under the imposed frame tension.

For rod moves with force or torque bias enabled, the contributions
from membrane vertices are
\begin{equation}
\mathbf F=\sum_i\mathbf F_i,\qquad
\boldsymbol{\tau}
=\sum_i(\mathbf q_i-\mathbf R_{\mathrm{rod}})\times\mathbf F_i,
\qquad
\mathbf F_i
=U_i'(r_i)\frac{\mathbf x_i-\mathbf q_i}{r_i}.
\end{equation}
Here $\mathbf q_i$ is the closest point on the rod-axis segment to
vertex $i$, $\mathbf R_{\mathrm{rod}}$ is the rod centre, and
$U_i(r)=s_i u_{\mathrm{att}}(r)+(1-s_i)u_{\mathrm{rep}}(r)$.
The derivative is evaluated within the selected branch, with $s_i$
held fixed.

When force bias is enabled, a trial translation combines a Gaussian
random displacement with a force-dependent drift whose magnitude
is capped by the translation-step parameter. Without force bias,
each displacement component is sampled uniformly within the
prescribed translation range. When torque bias is enabled, the
orientation is updated as
\begin{equation}
\hat{\mathbf u}'=
\frac{\hat{\mathbf u}+\delta\mathbf u}
     {|\hat{\mathbf u}+\delta\mathbf u|},
\end{equation}
where $\delta\mathbf u$ is perpendicular to the current rod axis and combines a Gaussian contribution with a torque-dependent drift
whose magnitude is capped by the rotation-step parameter. Without torque bias, a random rotation is proposed. Orientation changes are disabled for rods constrained to remain upright. Translation and orientation changes in a rod trial are accepted or rejected together.

Moves that violate geometric constraints are rejected. For symmetric proposals at fixed box size, the Metropolis acceptance probability is
\begin{equation}
P_{\mathrm{acc}}=\min\left[1,\exp(-\beta\Delta E)\right],
\end{equation}
where $\Delta E$ is the energy change included in the corresponding
acceptance test and $\beta=(k_{\mathrm B}T)^{-1}$.
In the soft-particle simulations, membrane-particle interaction energy changes are included in vertex-move acceptance tests but omitted from edge-flip acceptance tests. Biased rod moves  dditionally include forward and reverse Gaussian proposal corrections. Box-move acceptance includes the coordinate-rescaling Jacobian and the correction associated with the box-area sampling measure. We use reduced units with $k_{\mathrm B}T=1$.

\subsection{Simulation protocol}

The wrapping membrane is represented by a triangulated mesh containing 1968 vertices and 3936 triangles. The minimum allowed distance between membrane vertices, $\ell_0$, defines the unit of length, with $\ell_0=1$ in reduced units. Edge lengths are restricted to $[\ell_0,\sqrt{3}\ell_0]$. Starting from a planar configuration, we relax the membrane at zero imposed frame tension, allowing both its shape and projected area to fluctuate. The resulting nearly planar configuration is used to initialize the wrapping simulations, with box dimensions $L_x=L_y=54.5\,\ell_0$ and $L_z=100\,\ell_0$. The particle is then introduced as described below, and the frame tension is set to the prescribed value.

For a rigid spherical particle, a membrane vertex at $\mathbf x_0$ is selected at random, and the particle centre is placed at $\mathbf r_c=\mathbf x_0+R\hat{\mathbf n}_0$, where $\hat{\mathbf n}_0$ is the local unit normal. For a rod, its centre is placed at $\mathbf r_c=\mathbf x_0+(c+a/2)\hat{\mathbf n}_0$, and its initial axis is $\hat{\mathbf u}=\hat{\mathbf n}_0$. These placements put the selected vertex at the effective particle surface, with the rod initially aligned along the local membrane normal.

In the rod simulations, each MC sweep comprises 1968 membrane vertex-displacement attempts, 1180 edge-flip attempts, one combined rod translation and orientation attempt, and one isotropic box-rescaling attempt. The rod proposals use Gaussian translational increments with a standard deviation of $0.01\,\ell_0$ per Cartesian component and Gaussian orientation increments with a standard deviation of $0.01$ per tangent-plane component. The force- and torque-bias coefficients are both $10^{-4}$ in simulation units. Box moves propose a change in $L_x$ drawn uniformly from $[-0.01,0.01]\,\ell_0$, with $L_y$ rescaled by the same factor and $L_z$ held fixed.

The deformable particle is initialized from a separate closed mesh containing 650 vertices and 1296 triangles. Its initial position and shape are specified by the input vertex coordinates. The mesh is placed above the wrapping membrane, with a minimum distance of $1.1\,\ell_0$ between vertices belonging to the two surfaces. During the simulation, vertices on both surfaces can move, while edge flips are restricted to the wrapping membrane. Same as the vertex move in the planar membrane, each Cartesian component of a trial vertex displacement is drawn uniformly from $[-0.05,0.05]\,\ell_0$.

All simulation are run for $10^6$ MC sweeps. Configurations are saved every 1000 sweeps, and energy and box observables are recorded every 1000 sweeps.

\subsection{Calculation of the wrapping fraction}

We estimate the wrapping fraction $\alpha$ from the membrane configurations using a geometric estimator appropriate to each particle representation. Periodic images are taken into account when evaluating distances.

\paragraph{Spherical particle}

For a rigid sphere, we estimate coverage from the solid angles subtended by nearby membrane triangles at the particle centre. A membrane vertex is classified as close to the sphere when its radial distance satisfies $R-\delta\le r_i\le R+\delta$, where $\delta$ is the radial tolerance. A triangle is included if at least two of its three vertices satisfy this condition. The wrapping fraction is calculated as
\begin{equation}
    \alpha_{\mathrm{sphere}}
    =\min\left(1,\frac{1}{4\pi}\sum_{T\in\mathcal C}\Omega_T\right),
\end{equation}
where $\mathcal C$ is the set of selected membrane triangles and $\Omega_T$ is the unsigned solid angle subtended by triangle $T$ at the sphere centre. The radial tolerance is $\delta=1.5\,\ell_0$.

\paragraph{Rod}
Similar to the spherical particle case, rod wrapping is estimated by sampling points on its spherocylindrical surface, including the cylindrical region and both hemispherical caps. A point is classified as covered if its distance to the nearest membrane vertex, accounting for periodic boundaries in the $xy$ plane, is no greater than $1.5\,\ell_0$. The wrapping fraction is
\begin{equation}
    \alpha_{\mathrm{rod}}=\frac{\sum_j a_j I_j}{2\pi L r_{rod}+4\pi r_{rod}^2},
\end{equation}
where $a_j$ is the surface area represented by sampling point $j$, and $I_j$ equals one for a covered point and zero otherwise.

\paragraph{Deformable particle}

For a deformable particle, coverage is evaluated directly on its instantaneous triangular mesh. A particle vertex at $\mathbf y_j$ is classified as contacting the wrapping membrane if at least one membrane vertex at $\mathbf x_i$ satisfies
\begin{equation}
    |\mathbf d_{ij}|\le d_{\mathrm{contact}},
    \qquad
    \hat{\mathbf n}^{\mathrm p}_j\cdot\mathbf d_{ij}\ge0,
    \qquad
    \hat{\mathbf n}^{\mathrm m}_i\cdot\mathbf d_{ij}\le0,
\end{equation}
where $\mathbf d_{ij}=\mathbf x_i-\mathbf y_j$ is evaluated with periodic images, and $\hat{\mathbf n}^{\mathrm p}_j$ and $\hat{\mathbf n}^{\mathrm m}_i$ are the particle and membrane vertex normals. These normals are obtained by normalizing the sum of the oriented area vectors of adjacent triangles. The distance threshold is $d_{\mathrm{contact}}=1.5\,\ell_0$.

Let $I_j$ be one for a contacting particle vertex and zero otherwise. Each particle triangle contributes its area multiplied by the fraction of its vertices in contact:
\begin{equation}
    \alpha_{\mathrm{soft}}
    =\frac{\displaystyle\sum_{T\in\mathcal T_{\mathrm p}}
        A_T\left(\frac{1}{3}\sum_{j\in T}I_j\right)}
    {\displaystyle\sum_{T\in\mathcal T_{\mathrm p}}A_T},
\end{equation}
where $\mathcal T_{\mathrm p}$ contains the particle triangles and $A_T$ is the triangle area in the analyzed configuration.

For the rod and deformable-particle tension-adhesion scans, the reported wrapping fraction is evaluated from the last available saved configuration of each run.

\section{MesoMem}
\label{supp-app:MesoMem}
MesoMem is a particle based coarse-grain membrane model, particularly useful and efficient for modeling mesoscale systems such as particle wrapping or full vesicle deformation. To summarize the working principles: MesoMem represents a patch of lipid bilayer ($\approx5\times5$ nm) as beads that interact with each other based on potentials that invoke self-assembly of the particle into elastic sheets, while also allowing for in-plane fluidity. The parameters of the interactions can be tuned to match the characteristics of the membrane sheets to physiologically accurate values. Moreover, we can also generate various geometrical shapes such as spherical and tubular vesicles. For more details on the equations that govern the system and the validation of the model, refer to \cite{sillanoMesoMem2026}.

\subsection{Construction of the nanoparticles}
\paragraph{Adhesive potential}
In all cases, adhesion between the membrane and a nanoparticle is modeled with an appropriately cut-off Lennard-Jones potential: a short-range attraction, an excluded-volume core at short distance, and no interaction beyond the cutoff. For a nanoparticle at distance $d$ from a membrane particle,
\begin{equation}
U_{\text{LJ}}(d) = \begin{cases} 4w\left[\left(\dfrac{\ell_{\text{eff}}}{d}\right)^{12} - \left(\dfrac{\ell_{\text{eff}}}{d}\right)^{6}\right], & d \le r_c \\ 0, & d > r_c \end{cases}
    \label{supp-eq:lj_mesomem}
\end{equation}
where $w$ sets the adhesion strength, $\ell=1$ is the diameter of a membrane bead, $\ell_{\text{eff}} = \ell/2 + r_p$ is the sum of the membrane bead radius and the nanoparticle radius $r_p$, and $r_c = 2^{1/6}\ell_{\text{eff}} + \ell/2$ is the cut-off distance beyond which the interaction vanishes. What distinguishes the spherical and anisotropic cases below is only how $d$ and $r_p$ are defined, and whether the resulting force can exert a torque.

\paragraph{Wrapping of a spherical particle}
For a spherical nanoparticle of radius $r_p = r_{\text{pa}}$, both the particle and the membrane bead are point-like, so $d$ in Eq.~\eqref{supp-eq:lj_mesomem} is simply the center-to-center distance $r$ between them. The resulting force acts along the distance vector; since the interaction is isotropic and passes through the particle's center, no torque is generated.

\paragraph{Wrapping of an anisotropic particle}
To construct a spherocylindrical nanoparticle, we instead model it as a line segment of length $L$, with $r_p = r_{\text{rod}}$ the rod radius. Here $d$ is the distance from the membrane particle to the \emph{closest point on the rod's axis}, found by projecting the membrane particle onto the segment. Since the repulsive core sits at $\ell_{\text{eff}} = \ell/2 + r_{\text{rod}}$ around this entire axis, the particle effectively becomes a solid rod of thickness $2r_{\text{rod}}$ and total length $h = L + 2r_{\text{rod}}$ (the axis length plus two hemispherical caps of radius $r_{\text{rod}}$) — i.e., a spherocylinder.

Because the rod is not point-like, the force still acts along the vector between the membrane particle and its contact point $C$ on the axis, but $C$ generally does not coincide with the rod's center $\mathbf{x}_{\text{rod}}$. Combined with the rod's orientational degree of freedom, this lever arm produces a torque,
\begin{equation}
    \boldsymbol{\tau} = (\mathbf{C} - \mathbf{x}_{\text{rod}}) \times \mathbf{F},
    \label{supp-eq:torque_mesomem}
\end{equation}
which is what allows the membrane to reorient the rod as it wraps around it, rather than only displacing it translationally.

The membrane--particle interaction described above is implemented as an additional custom pair potential in LAMMPS \cite{LAMMPS}, which handles both the isotropic and the anisotropic case and, for the latter, the associated torque. It is provided, together with the rest of the simulation and analysis code, in our GitLab repository \url{https://gitlab.tudelft.nl/idema-group/MesoMem}.

\subsection{Simulation protocol}
We generate a planar sheet made of $N=5480$ membrane particles, and position the wrapping particle above the membrane. The protocol is described here for a rod; the wrapping of a spherical particle follows exactly the same four phases, the only difference being the particle geometry, as a sphere has no orientational degree of freedom and therefore no freely-rotating, rocket-like or submarine-like variants. The rod's initial alignment is perpendicular to the membrane for the freely-rotating and rocket-like cases, whereas for the submarine-like case it is parallel to it. Initially, the length of one side of the box is set to be approximately $62.5\ell$, such that the membrane beads can fill the x-y plane in the middle of the box without any wide holes. One simulation run is composed of 4 phases:
\begin{itemize}
    \item \textbf{Phase 1}: The membrane is relaxed in NVT ensemble (using a Langevin thermostat \cite{LAMMPS_fix_langevin}) without interacting with the rod. 
    \item \textbf{Phase 2}: The system switches from the NVT ensemble to a fixed lateral tension ensemble. In this ensemble, the integration is performed using a Nose/Hoover barostat, where the box size is allowed to change in the x- and y-directions so that a target in-plane pressure ($P_{xx}, P_{yy}$) is maintained. The lateral pressure, then, can be converted to lateral tension, $\tau$ according to \cite{ghoufi_mesoscale_2011,kirkwood_statistical_1949}:
    \begin{equation}
        \tau = \int \left[ P_N(z) - P_T(z) \right] dz = L_z \left( P_{zz} - \frac{P_{xx} + P_{yy}}{2} \right).
    \end{equation}
    During this phase the target in-plane pressure is set to zero, so that the membrane relaxes at vanishing lateral tension. 
    \item \textbf{Phase 3}: The system begins equilibrating towards the target value defined for the lateral pressure. The value of the pressure is set to be negative so the box size increases in the x-y plane. This ensures that a stretching tension is applied to the membrane sheet.
    \item \textbf{Phase 4}: The rod is allowed to move and the interaction between the rod and the membrane is turned on. In freely-rotating simulations, the rod is allowed to experience torque, whereas in the other workflows its direction is fixed. The rod's motion is integrated with \texttt{nve/sphere} and coupled to the \texttt{langevin} thermostat. The rod is initialized with an initial small velocity towards the membrane to facilitate contact.
\end{itemize}
Simulations are halted and removed from further analysis in two situations:
\begin{enumerate}
    \item if the membrane rips due to overstretching, which we detect when the box expands beyond a limit value;
    \item if the rod never comes into contact with the membrane. 
\end{enumerate}

\subsubsection*{Calculation of the wrapping fraction}
For a given timestep, the wrapping fraction $\alpha$ is estimated based on the following algorithm:
\begin{itemize}
    \item The surface of the rod is divided into uniformly distributed points.
    \item For each surface point, distance to its closest membrane bead is calculated. If the value is below a threshold, that surface point is considered to be covered.
    \item The minimum of the Lennard-Jones potential is at the distance
    \begin{equation}
        r_{\text{min}}=1.112\ell_{\text{eff}},
    \end{equation}
    which means that the vertical distance of the bead from the surface of the rod becomes $r_{\text{min}}-r_{\text{rod}}=1.112r_{\text{bead}}+0.112r_{\text{rod}}$. Hence, the distance is dependent on the radius of the rod. Considering that the thermal fluctuations will also cause the distance of bound beads to vary, choosing a fixed contact threshold for all rods can lead to mistakes in the counting of bound beads.
    
    Therefore, the contact threshold is determined through a separate algorithm that analyzes snapshots on a case-by-case basis. First, it takes the membrane beads whose projection on the rod's axis line sits between cylindrical part of the rod $[-L/2, L/2]$. Then it measures the vertical distances of these membrane beads to the axis line of the rod. In this list of distance values, the binding distance is calculated as the value below which 5 percent of the smallest distance values lie. Hence, we get a good estimate of the distance that the closest (i.e. bound) beads have to the rod surface. We also add a small tolerance to the binding distance to get the contact threshold, so that we can account for the area that the cross-section of the beads cover.
    \item Then, the wrapping fraction for each frame is calculated as the ratio of covered surface points to the total amount.
    \item Lastly, we average over the frames after the wrapping-fraction time series has reached a plateau. Each run is repeated 3 times with the same set of parameters.
    \item Once full wrapping is achieved, we observe that the wrapped capsule detaches from the membrane, so the wrapping fraction does not decrease again after the plateau.
\end{itemize}

\section{Influence of spherical particle size on wrapping}
\label{supp-app:particle_size}

\begin{figure}[ht]
    \centering
    \includegraphics[width=\textwidth]{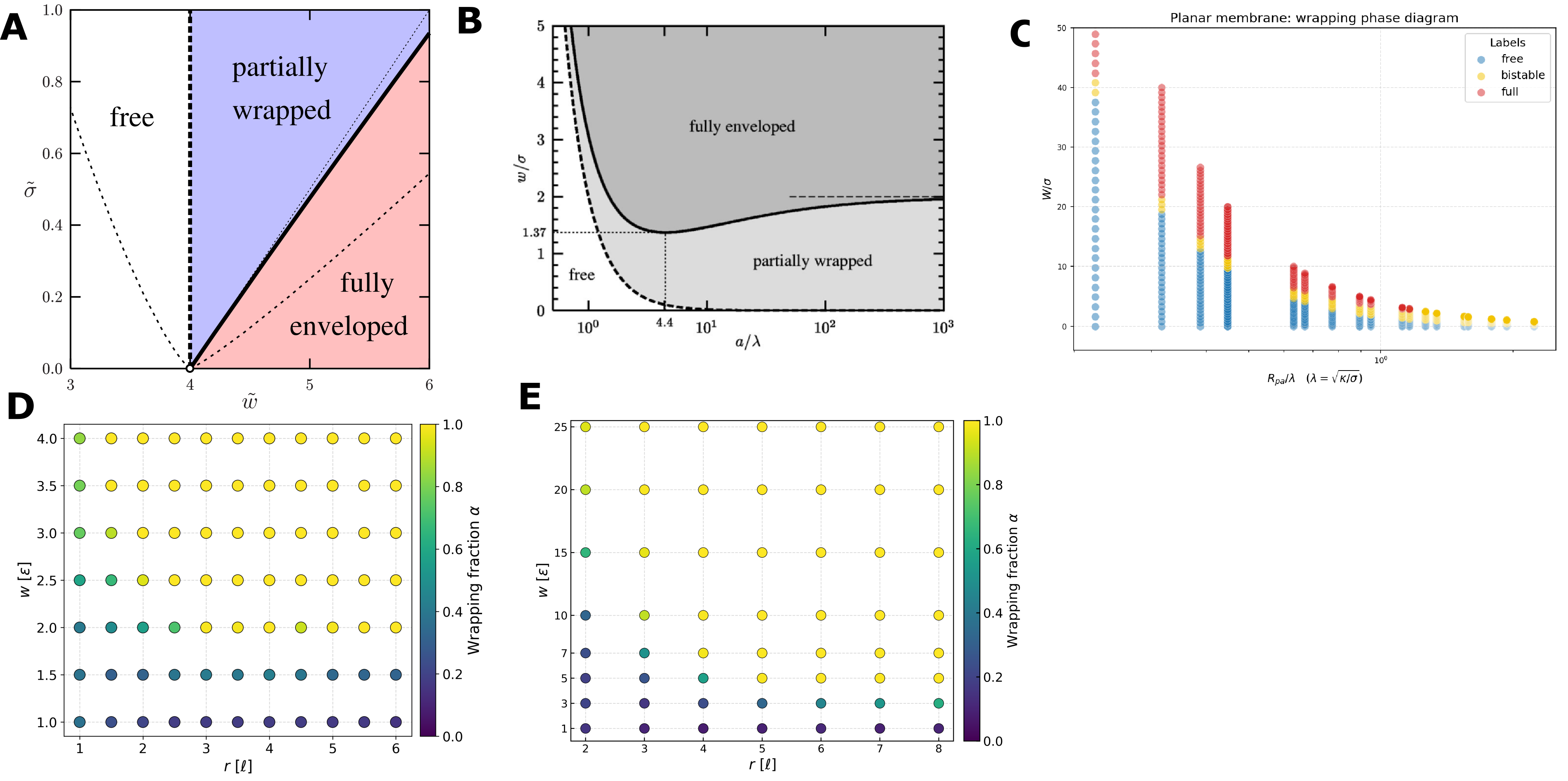}
    \caption{(A, B) Phase diagrams adapted from figure 2 and figure 5 respectively of \cite{desernoElasticDeformationFluid2004} reproduced with permissions, in the rescaled adhesion--tension plane (A) and as a function of the reduced particle size $a/\lambda$ (B). (C) The same phase diagram as (B) obtained with our implementation of the planar solver, with each point classified as free (blue), bistable (yellow), or fully wrapped (red). (D) MesoMem and (E) FreeDTS simulations, in which the colour denotes the equilibrium wrapping fraction; each MesoMem point is an average over 3 independent replicas at zero lateral tension, and each FreeDTS point is a single simulation at a zero lateral tension.}
    \label{supp-fig:rpa_plot_comparison}
\end{figure}

Particle size enters the wrapping problem through the ratio between the particle radius $R_{\mathrm{pa}}$ and the membrane relaxation length $\lambda=\sqrt{\kappa/\sigma}$, which sets how far the deformation induced by the particle extends into the surrounding membrane. Fig.~\ref{supp-fig:rpa_plot_comparison} collects this size dependence for a spherical particle on a planar membrane, comparing the published reference result with the three approaches used in this work. Panels (A) and (B) reproduce the phase diagrams of Deserno \cite{desernoElasticDeformationFluid2004}, in the rescaled adhesion--tension plane and as a function of the reduced particle size $R_{\mathrm{pa}}/\lambda$, respectively. Panel (C) is the same diagram computed with our implementation of the planar solver, each point being classified as free, bistable, or fully wrapped from the shape of its energy landscape (Sec.~\ref{supp-sec:classifying_states}). Panels (D) and (E) report the equilibrium wrapping fraction measured over the same range of particle radii in MesoMem and FreeDTS simulations.

Two trends are visible across the three methods. The boundary between the free and the partially wrapped state moves to weaker adhesion as the particle grows, so that large particles bind more easily, whereas the boundary to complete wrapping is non-monotonic: it passes through a minimum at $R_{\mathrm{pa}}/\lambda\approx4.4$, so that both small and large particles are fully engulfed at lower adhesion strength than intermediate ones. We use this survey mainly to fix the particle radius used in the simulations, choosing a value for which the free, partially wrapped, and fully wrapped states are all accessible.

\section{Validation of the vesicle solver}
\label{supp-app:vesicle_validation}

\begin{figure}[ht]
    \centering
    \includegraphics[width=\textwidth]{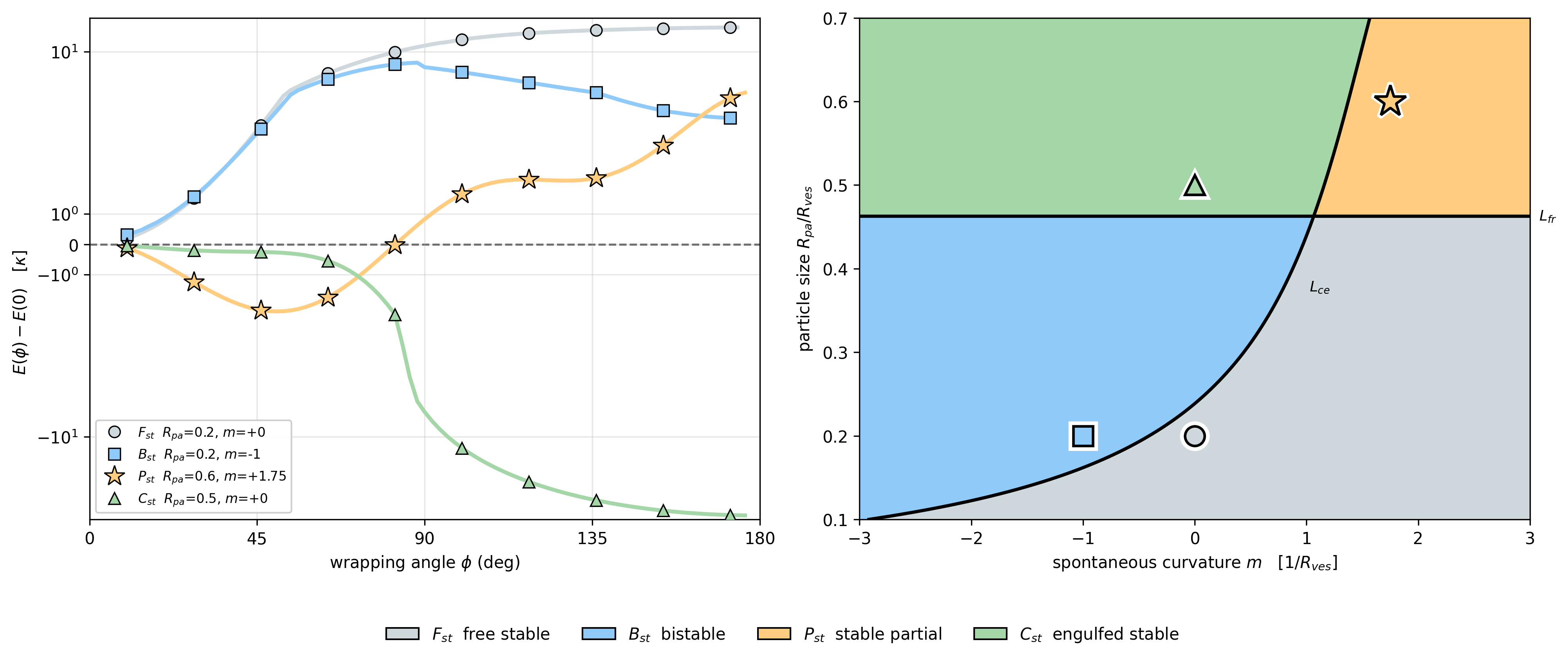}
    \caption{Validation of the closed-vesicle solver against the four wrapping states of Agudo-Canalejo and Lipowsky \cite{agudo-canalejoCriticalParticleSizes2015}, for a rigid spherical particle of radius $R_{\mathrm{pa}}$ adhering to the outside of a vesicle of radius $R_{\mathrm{ves}}$.
    (Left) Free-energy landscapes $E(\phi)-E(0)$, in units of the bending rigidity $\kappa$ and plotted on a symmetric logarithmic scale, as a function of the wrapping angle $\phi$. Each curve is computed at the particle size and spontaneous curvature given in the legend and illustrates one of the four states.
    (Right) State diagram in the plane of spontaneous curvature $m$, in units of $1/R_{\mathrm{ves}}$, and relative particle size $R_{\mathrm{pa}}/R_{\mathrm{ves}}$: $F_{\mathrm{st}}$, free state stable (grey); $B_{\mathrm{st}}$, bistable (blue); $P_{\mathrm{st}}$, partially wrapped state stable (orange); and $C_{\mathrm{st}}$, completely engulfed state stable (green). The solid curves $L_{\mathrm{fr}}$ and $L_{\mathrm{ce}}$ are the stability limits of the free and of the completely engulfed state, respectively. The four markers indicate the parameter pairs at which the landscapes on the left were computed.}
    \label{supp-fig:vesicle_validation}
\end{figure}

We validate the closed-vesicle solver by reproducing the four wrapping states identified by Agudo-Canalejo and Lipowsky \cite{agudo-canalejoCriticalParticleSizes2015} for a rigid spherical particle adhering to the outer surface of a vesicle. Sweeping the spontaneous curvature $m$ and the relative particle size $R_{\mathrm{pa}}/R_{\mathrm{ves}}$ recovers their state diagram, including the stability limits $L_{\mathrm{fr}}$ and $L_{\mathrm{ce}}$ of the free and of the completely engulfed state (Fig.~\ref{supp-fig:vesicle_validation}).

\section{Landscapes for varying aspect ratios}
\label{supp-app:aspect_ratio_landscapes}

\begin{figure}[ht]
    \centering
    \includegraphics[width=\textwidth]{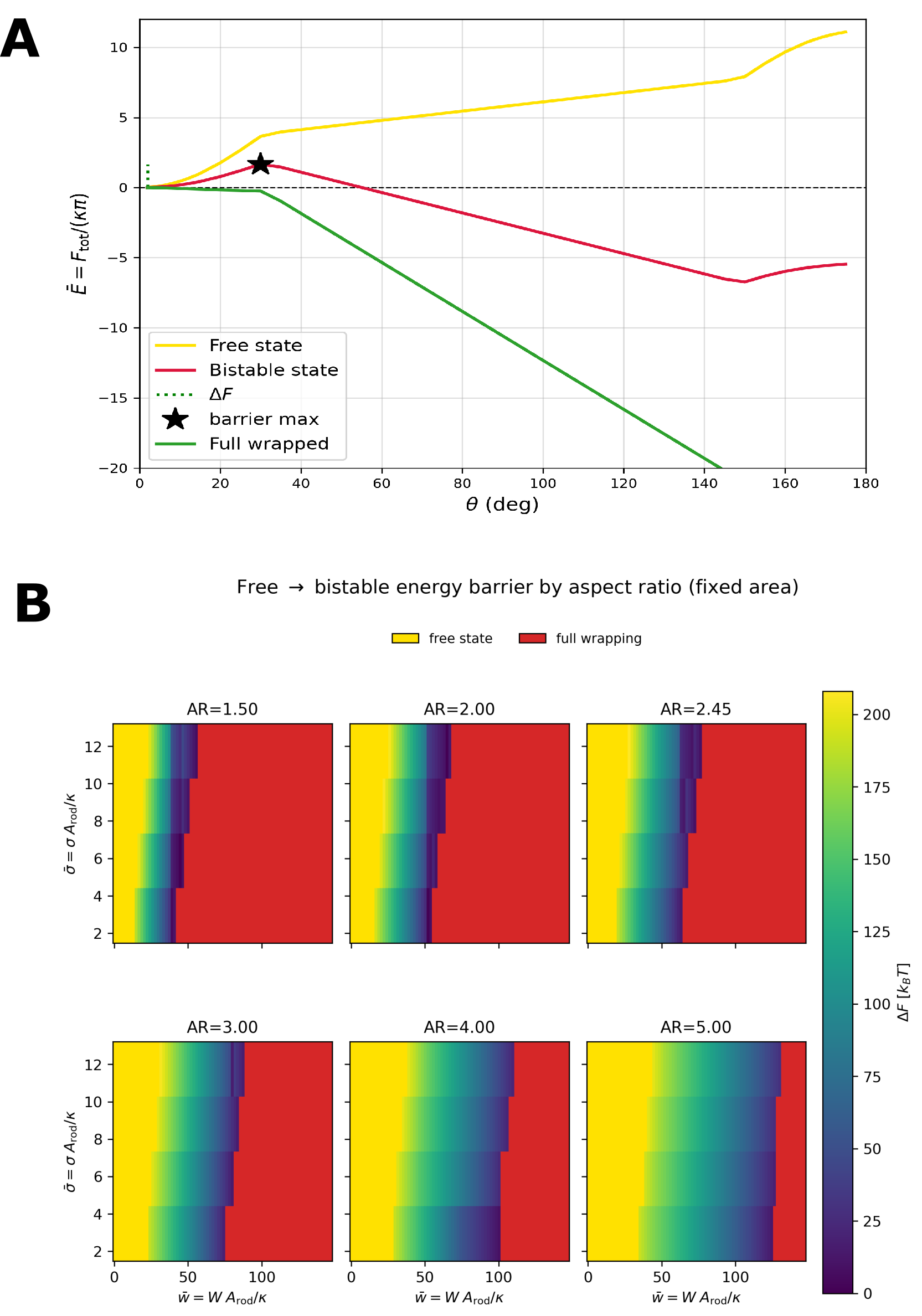}
    \caption{(A) Representative free-energy landscapes of the free (yellow), bistable (red), and fully wrapped (green) types, the star marking the barrier maximum.  (B) Influence of the particle aspect ratio $AR$ on the wrapping behaviour of a spherocylindrical particle. The colour coding is the same as in the main-text figure, Fig.~\ref{fig:rod_aspect_ratio}.}
    \label{supp-fig:aspect_ratio_landscapes_supp}
\end{figure}

The main text characterises the wrapping of spherocylindrical rods through the energy barriers of Fig.~\ref{fig:rod_aspect_ratio}C--D; here we show the free-energy landscapes those barriers are extracted from. Fig.~\ref{supp-fig:aspect_ratio_landscapes_supp}(A) shows one representative landscape of each of the three types encountered for rods of fixed surface area: a monotonically increasing landscape, for which the free state is the only minimum; a monotonically decreasing one, for which the particle is completely wrapped; and a bistable landscape, in which a partially wrapped minimum is separated from the free state by a barrier, whose maximum is marked by a star. Panel (B) shows the resulting classification for the six aspect ratios between $1.5$ and $5.0$ considered in the main text, in the plane of the rescaled adhesion strength $\bar w=WA_{\mathrm{rod}}/\kappa$ and rescaled lateral tension $\bar\sigma=\sigma A_{\mathrm{rod}}/\kappa$, with the barrier height colour-coded inside the bistable region.

Increasing the aspect ratio at fixed particle area shifts both boundaries towards stronger adhesion and widens the bistable band, in agreement with the main text. The magenta points, where the barrier drops below $10\,k_{\mathrm{B}}T$ and the transition to the wrapped state can be crossed by thermal fluctuations, are found only for the more compact shapes and essentially disappear for the most elongated ones.

\section{Numerical case for ellipsoidal particles}
\label{supp-app:ellipsoids}
\begin{figure}[ht]
    \centering
    \includegraphics[width=\textwidth]{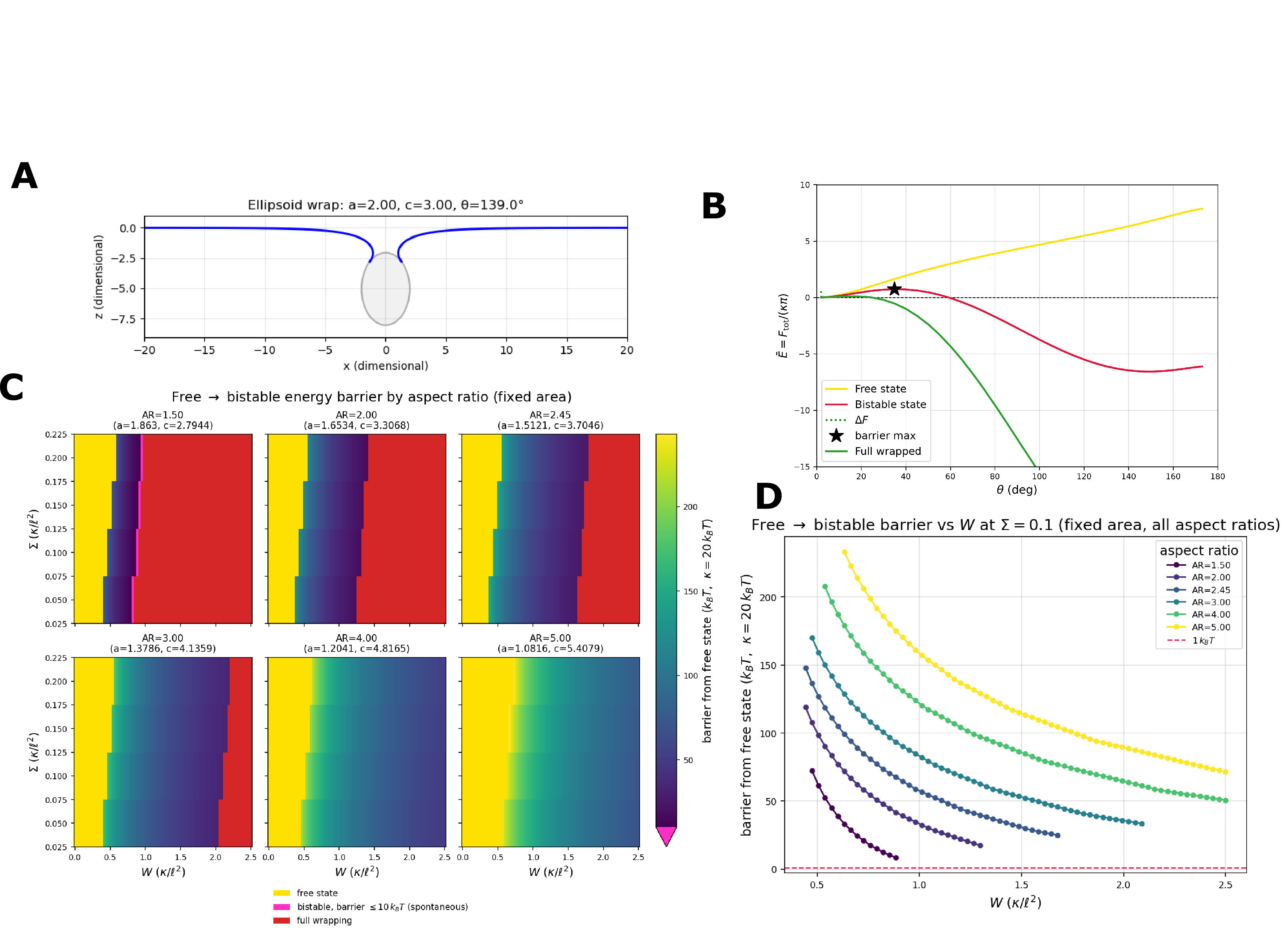}
    \caption{Influence of the particle aspect ratio on the wrapping behaviour of ellipsoidal particles of equal surface area, obtained with the planar numerical solver. (A) Membrane profile around an ellipsoid with semi-axes $a=2$, $c=3$ at a wrapping angle of $139^\circ$. (B) Representative free-energy landscapes of the free (yellow), bistable (red), and fully wrapped (green) types, the star marking the barrier maximum. (C) State diagrams in the plane of adhesion strength $W$ and lateral tension $\Sigma$ for six aspect ratios between $1.5$ and $5.0$: yellow marks the free state, red the fully wrapped state, the viridis gradient the barrier height inside the bistable region. (D) Barrier height as a function of $W$ at fixed $\Sigma=0.1\,\kappa/\ell^2$ for each aspect ratio; the dashed red line marks the thermal energy $k_{\mathrm{B}}T$.}
    \label{supp-fig:ellips_aspect_ratio}
\end{figure}

As mentioned in the main text, we considered not only spherocylindrical particles but also ellipsoidal ones. In Fig.~\ref{supp-fig:ellips_aspect_ratio} we repeat for prolate ellipsoids of equal surface area the analysis carried out for rods, the particle long axis being aligned with the membrane normal as required by axial symmetry.

The behaviour parallels that of rods: elongation displaces both boundaries towards stronger adhesion and raises the barrier separating the free from the partially wrapped state. The barriers, however, remain of the order of tens to hundreds of $k_{\mathrm{B}}T$ over the whole range explored, far above the thermal energy.

\section{Soft-particle volume and wrapping geometry}

\begin{figure}[ht]
    \centering
    \includegraphics[width=0.95\textwidth]{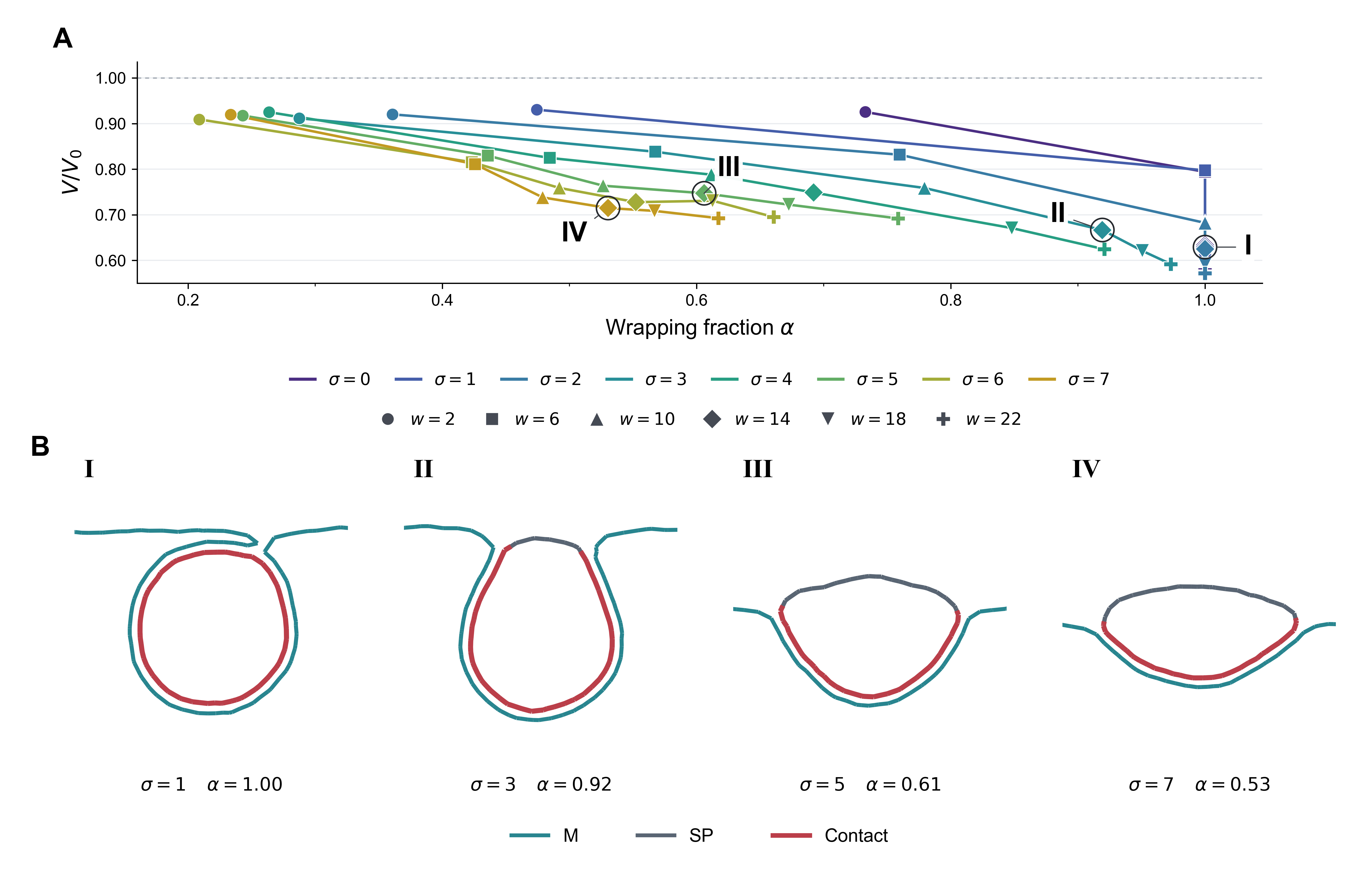}
    \caption{(A) Relative particle volume $V/V_0$ versus wrapping fraction $\alpha$ after $10^6$ MC steps, where $V_0$ is the initial soft particle volume. Colors indicate membrane frame tension $\sigma$, and marker shapes indicate adhesion strength $w$. Lines connect independent simulations at fixed $\sigma$. (B) Cross-sectional configurations for the four circled states I-IV at $w=14\,k_{\mathrm B}T$ and $\sigma=1,3,5,7\,k_{\mathrm B}T/\ell_0^2$, respectively. Teal, grey, and red denote the membrane (M), soft particle (SP), and contact region, respectively.}
    \label{supp-fig:soft_V}
\end{figure}

At fixed membrane tension, stronger adhesion generally increases the wrapping fraction while reducing the particle volume (Fig.~\ref{supp-fig:soft_V}A). At low tensions, the relative volume continues to vary even among states with $\alpha \simeq 1$, showing that full surface coverage does not uniquely determine particle volume.

The sections at $w=14\,k_{\mathrm B}T$ show how the geometry changes with tension (Fig.~\ref{supp-fig:soft_V}B, I-IV). At $\sigma=1$, the particle has a nearly round cross-section and a narrow membrane neck. At $\sigma=3$, it remains deeply wrapped but has a more elongated cross-section. At $\sigma=5$ and $7$, wrapping is shallower and the sections are broader and flatter. Configurations I and II have comparable relative volumes despite their different outlines and contact regions. Thus, neither wrapping fraction nor particle volume alone describes the final geometry.

\printbibliography

\end{refsection}

\end{document}